\documentclass[letterpaper]{JHEP3}
\usepackage{amsfonts}

\newcommand{\ab}{{(j)}}
\newcommand{\bb}{{(k)}}

\def\inbar{\,\vrule height1.5ex width.4pt depth0pt}
\def\I1{\relax\hbox{$\inbar\kern-.35em{\rm 1}$}}
 \font\cmss=cmss10 \font\cmsss=cmss10 at 7pt

\def\IZ{\relax\ifmmode\mathchoice
 {\hbox{\cmss Z\kern-.4em Z}}{\hbox{\cmss Z\kern-.4em Z}}
 {\lower.9pt\hbox{\cmsss Z\kern-.4em Z}}
 {\lower1.2pt\hbox{\cmsss Z\kern-.4em Z}}
 \else{\cmss Z\kern-.4em Z}\fi}

\newcommand{\vphi}{\varphi}

\newcommand{\eq}{\begin{eqnarray}}
\newcommand{\feq}{\end{eqnarray}}
\newcommand{\eqn}{\begin{eqnarray}}
\newcommand{\feqn}{\end{eqnarray}}
\newcommand{\arr}{\begin{eqnarray*}}
\newcommand{\farr}{\end{eqnarray*}}

\newcommand{\p}{\partial}

\usepackage{epsfig}

\abstract{
We explore via linearized perturbation theory 
the Gregory-Laflamme instability of rotating black strings
with equal magnitude angular momenta.
Our results indicate that the Gregory-Laflamme 
instability persists up to extremality
for all even dimensions between six and fourteen.
We construct rotating nonuniform black strings with 
two equal magnitude angular momenta in six dimensions.
We see a first indication
for the occurrence of a topology changing transition,
associated with such rotating nonuniform black strings.
Charged nonuniform black string
configurations 
in heterotic string theory
are also constructed
by employing a solution generation technique.
}

\title{Rotating nonuniform black string solutions }

\author{Burkhard Kleihaus$^{1}$\thanks{%
E-mail: \texttt{kleihaus@theorie.physik.uni-oldenburg.de}}, \ Jutta Kunz$^{1}$%
\thanks{%
E-mail: \texttt{kunz@theorie.physik.uni-oldenburg.de}} \ and Eugen Radu$^2$\thanks{%
E-mail: \texttt{radu@thphys.nuim.ie}} \\
$^{1}$Institut f\"ur Physik, Universit\"at Oldenburg, Postfach 2503
D-26111 Oldenburg, Germany\\ 
$^{2} $Department of Mathematical Physics, National University of Ireland,
Maynooth, Ireland}

\begin{document}

\section{Introduction}  
In recent years interest in the properties of gravity in more than $D=4$ 
dimensions increased significantly.
This interest was enhanced by the development of string theory, 
which requires a ten-dimensional spacetime,
to be consistent from a quantum point of view.
In order not to contradict observational evidence, 
the extra dimensions are usually supposed to be compactified
on small scales. 

Black string solutions, present for
$D\geq 5$ spacetime dimensions, are of particular interest, since they 
exhibit new features that have no analogue in the black hole case.
The simplest vacuum static solutions of this type 
are found by trivially extending to $D$ dimensions  
the vacuum black hole solutions to Einstein equations in $D-1$ dimensions.
These then correspond to uniform black strings (UBS)
with horizon topology $S^{D-3}\times S^1$.   

One of the first steps towards understanding the 
higher-dimensional solutions is to investigate
their classical stability against small perturbations.
In a surprising development, Gregory and Laflamme (GL) showed that
the static UBS solutions are unstable below a critical value of the mass
\cite{Gregory:1993vy}. 
Following this discovery, a branch of nonuniform black string (NUBS)
solutions was found perturbatively from the critical
GL string in five \cite{Gubser:2001ac}, six \cite{Wiseman:2002zc} 
and in higher, up to sixteen, dimensions \cite{Sorkin:2004qq}. 
This nonuniform branch was numerically extended into the full nonlinear regime
for $D=5$ \cite{Kleihaus:2006ee} and $D=6$ \cite{Wiseman:2002zc,Kleihaus:2006ee}.
In recent work \cite{Sorkin:2006wp}
the nonuniform branch was extended for all dimensions up to eleven.
These NUBS are static configurations with a nontrivial dependence 
on the extra dimension, 
and their mass is always greater than that of the critical UBS
(see \cite{Kol:2004ww} and also \cite{Harmark:2007md} 
for a recent review of this subject).

Apart from the black string solutions, 
Kaluza-Klein (KK) theory possesses also a 
branch of black hole solutions with an event horizon of topology $S^{D-2}$.
The numerical results presented in \cite{Kudoh:2004hs, Kleihaus:2006ee}
(following a conjecture put forward in \cite{Kol:2002xz})
suggest that the black hole branch and the nonuniform
string branch merge at a topology changing transition.
The problem is also interesting as it is connected 
by holography to the phase structure of large $N_c$
super-Yang-Mills theory at strong 
coupling, compactified on a circle (see e.g. \cite{Harmark:2004ws},
\cite{Harmark:2006df},
\cite{Chowdhury:2006qn} ).

Recently, interest in the properties of rotating solutions
in more than $D=4$ dimensions increased significantly, as well.
Rotating black objects typically exhibit much richer dynamics than their
static counterparts, especially in more than four dimensions.
A famous example is the black ring solution \cite{Emparan:2001wn} 
in five-dimensional vacuum gravity,
which has horizon topology $S^2\times S^1$, its tension
and self gravitational attraction being balanced by the rotation of the ring.

The KK theory presents also spinning configurations.
The simplest rotating UBS configurations are found by taking 
the direct product of a $(D-1)$-dimensional Myers-Perry (MP) solution
\cite{Myers:1986un}
with a circle. 
These solutions are likely to exhibit a
classical GL instability, as well, at least for some range of the parameters.
However, previous investigations have focused on static black strings,
and no attention has been given to spin.
A major obstacle in this direction
is that the analytic theory of perturbations of higher-dimensional black holes
has not been fully developed yet
(see, however, the recent work \cite{Kunduri:2006qa}).

A MP spinning
black string in $D$ dimensions is characterized by 
the mass-energy, the tension, and 
$[(D-2)/2]$ angular momenta,
where $[(D-2)/2]$ denotes the integer part of $(D-2)/2$.
The generic rotating nonuniform solutions possess a nontrivial dependence
on $(D-4)/2$ angular coordinates, 
which therefore pose a difficult numerical problem.
In the even-dimensional case, however,
the problem can be greatly simplified, when the
${\it a\ priori}$ independent $(D-2)/2$ angular momenta 
are chosen to have equal magnitude,
since this factorizes the angular dependence \cite{Kunz:2006eh}.
The problem then reduces to studying the solutions of a set of five partial
differential equations with dependence only
on the radial variable $r$ and the extra dimension $z$.

In this paper we focus on this particular case, by
studying first the GL instability of MP UBSs with equal 
magnitude angular momenta in even spacetime dimensions.
These particular MP solutions possess interesting features,
which strongly contrast with those of MP solutions
with a single nonzero angular momentum.
In particular, we note the existence of an upper bound for
the scaled angular momenta $J/M^{(D-3)/(D-4)}$ \cite{Myers:1986un,Kunz:2006jd},
an extremal solution being found,
when this limit is approached;
whereas no such upper bound is present
for single angular momentum MP solutions, unless $D=4+1$ or $D=5+1$.

Although the GL instability is inevitable for static vacuum 
black strings,
the presence of rotation (or a gauge field
charge) might prevent black strings from exhibiting
such an instability.
However, our numerical results indicate that the GL instability 
persists for rotating vacuum black strings all the way to extremality, 
at least for all (even) dimensions between six and fourteen.

This type of solutions also provides a new laboratory
to test the Gubser-Mitra (GM) conjecture \cite{Gubser:2000ec}, 
that correlates
the dynamical and thermodynamical stability 
for systems with translational symmetry and infinite extent.
In this conjecture, the appearance of a negative specific heat
of a black string is related to the onset of a classical instability
(see \cite{Miyamoto:2006nd, Kudoh:2005hf} for a discussion of this issue in the case 
of charged static black strings and black $p-$branes).
The analysis of the thermodynamical stability of the MP UBS 
indicates that thermodynamical
stability becomes possible when taking a canonical ensemble, for solutions
near extremality.
 However, in a grand canonical ensemble
all UBS configurations are unstable,
which agrees with the results we found when studying the 
GL instability.

In $D=6$ we constructed the set of rotating
nonuniform black strings numerically.
These rotating NUBS solutions can be found by
starting with static NUBS configurations
and increasing the value of angular velocity of the event horizon,
in the domain where static NUBS exist.
Alternatively, one can start with rotating MP UBS solutions
and then construct the set of rotating NUBS configurations
from the stationary perturbative nonuniform solutions.

The paper is structured as follows:
we begin with a presentation of the general ansatz and the generic
properties of
rotating black strings with equal angular momenta in  
even spacetime dimensions. 
In  Section 3, we consider the corresponding
 MP uniform solutions, discussing their
thermodynamical properties and the issue of GL instability. 
 In Section 4 we  
 demonstrate that for $D=6$
a set of rotating NUBS solutions with two equal angular momenta exists,
at least within the scope of our numerical approximation.
 The numerical methods used here are similar to those
 employed to obtain the $D=5,~6$ static NUBS solutions in \cite{Kleihaus:2006ee}.
We discuss charged NUBS in heterotic string theory in section 5,
and give our conclusions and remarks in the final section.

\section{General ansatz and properties of the solutions}  

\subsection{The equations } 
 
We consider the Einstein action
\begin{eqnarray} 
\label{action-grav} 
I=\frac{1}{16 \pi G_D}\int_M~d^Dx \sqrt{-g} R
-\frac{1}{8\pi G_D}\int_{\partial M} d^{D-1}x\sqrt{-h}K,
\end{eqnarray}
in a $D-$dimensional spacetime, with $D\geq 6$ an even number.
The last term in  (\ref{action-grav}) is the Hawking-Gibbons surface term \cite{Gibbons:1976ue},
which
is required in order to have a well-defined variational principle. 
$K$ is the trace 
of the extrinsic curvature for the boundary $\partial\mathcal{M}$ and $h$ is the induced 
metric of the boundary.  
 
We consider black string solutions approaching asymptotically the 
$(D-1)$-dimensional Minkowski-space times a circle ${\cal M}^{D-1}\times S^1$.
We denote the compact direction as $z = x^{D-1}$ 
and the directions of $R^{D-2}$ as $x^1,...,x^{D-2}$, while $x^D=t$.
The direction $z$ is periodic with period $L$.
We also define the radial coordinate $r$ by
$r^2 = (x^1)^2 + \cdots + (x^{D-2})^2$.

To obtain nonuniform generalizations of the rotating uniform black
string MP solutions,
we consider space-times with  $ (D-2)/2$ commuting Killing vectors
$ \partial_{\varphi_k}$.
While the general configuration will then possess $ (D-2)/2 $ independent
angular momenta, we here restrict to rotating NUBS whose
angular momenta have all equal magnitude.
Analogous to the case of black holes \cite{Kunz:2006eh},
the metric parametrization then simplifies considerably
for such rotating NUBS 
\begin{eqnarray}
&&ds^2 = -e^{2A(r,z)}\frac{ f(r)}{h(r)}dt^2 + e^{2B(r,z)} (\frac{ dr^2}{f(r)} +dz^2)
+
e^{2C(r,z)}r^2\sum_{i=1}^{(D-4)/2}
  \left(\prod_{j=0}^{i-1} \cos^2\theta_j \right) d\theta_i^2
\nonumber
 \\
\nonumber
&&
+e^{2G(r,z)}h(r)r^2 \sum_{k=1}^{(D-2)/2} \left( \prod_{l=0}^{k-1} \cos^2 \theta_l
  \right) \sin^2\theta_k \left( d\vphi_k - W(r,z)dt\right)^2+r^2\left(e^{2C(r,z)}-e^{2G(r,z)} h(r)\right)
\\
\label{metric}
&&
\times
\left\{ 
\sum_{k=1}^{(D-2)/2} \left( \prod_{l=0}^{k-1} \cos^2
 \theta_l \right) \sin^2\theta_k  d\vphi_k^2 \right.
  -\left. \left[\sum_{k=1}^{(D-2)/2} \left( \prod_{l=0}^{k-1} \cos^2
  \theta_l \right) \sin^2\theta_k   d\vphi_k\right]^2 \right\} \ ,
\end{eqnarray}
where  $\theta_i \in [0,\pi/2]$
for $i=1,\dots , (D-4)/2$,  $\vphi_k \in [0,2\pi]$ for $k=1,\dots , (D-2)/2$,
and we formally define $\theta_0 \equiv 0$,
$\theta_{(D-2)/2} \equiv \pi/2$.
As a result of taking all angular momenta to be equal,
the symmetry group of this spacetime is enhanced from $R\times U(1)^{(D-2)/2}$
to $R\times U(\frac{D-2}{2})$, where $R$ denotes time translation.

We shall assume that the information on the NUBS
solutions is encoded in the functions $A(r,z),B(r,z),C(r,z),G(r,z)$
and $W(r,z)$, while
  $f(r)$ and $h(r)$ are two `background' functions which are chosen
for convenience. 
A useful parametrization when studying unstable modes around a MP solution  is
\begin{eqnarray}
\label{MP}
 f(r)=1
-\frac{2M }{r^{D-4}}
+\frac{2Ma^2}{r^{D-2}}~,~~
h(r)= 1+\frac{2Ma^2}{r^{D-2}}~,~~
w(r)=\frac{2Ma}{r^{D-2}h(r)}~,
\end{eqnarray}
where $M$ and $a$ are two constants related to the solution's mass and 
angular momentum 
(and $W(r,z)=w(r)$ for uniform black strings).
When constructing nonperturbative NUBS solutions, a more convenient choice
for the numerics is
\begin{eqnarray}
\label{MP-2}
f(r)=1-(r_0/r)^{D-4}~,~~h(r)=1 ~,
\end{eqnarray} 
together with a redefinition of the radial coordinate,
where $r_0$ denotes the coordinate value of the horizon.
The static NUBS ansatz used in previous studies is recovered for
$G(r,z)=C(r,z)$ and $W(r,z)=0$.
 
A suitable combination 
of the Einstein equations,
 $G_t^t=0,~G_r^r+G_z^z=0$, $G_{\theta_1}^{\theta_1}=0$,
 $G_{\varphi_1}^t=0$  and  $G_{\varphi_1}^{\varphi_1}=0$,
yields the following set of equations for the functions $A,~B,~C,~G$ and $W$ 
\begin{eqnarray}
\label{ec1} 
&&\hat O^2 A
+(\hat O A)^2
+\hat O A \cdot \hat O G
+(D-4) \hat O A \cdot \hat O C
-\frac{e^{-2A+2G}h^2}{2r^2f}(\hat O W)^2
\\
\nonumber
&&
+(\frac{D-3}{r}+\frac{3f'}{2f}-\frac{h'}{2h})\partial_r A
+( \frac{f'}{2f}-\frac{h'}{2h})((D-4)\partial_r C+\partial_rG)
\\
\nonumber
&&
+\frac{(D-3)f'}{2rf}-\frac{(D-3)h'}{2rh}
-\frac{f'h'}{2fh}
+\frac{h'^2}{2h^2}
+\frac{f''}{2f}
-\frac{h''}{2h}=0,
\end{eqnarray}
\begin{eqnarray}
\label{ec2}
&&
 \hat O^2 B
-\frac{e^{-2A+2G}r^2h^2}{4f}(\hat O W)^2
-(D-4)\hat O A \cdot \hat O C
-\frac{1}{2}(D-4)(D-5)(\hat O C)^2
\\
\nonumber
&&
-\hat O A \cdot \hat O G
-(D-4)\hat O C \cdot \hat O G
-(\frac{D-3}{r}+\frac{h'}{2h})\partial_r A
+\frac{f'}{2f}\partial_rB
-(D-4)(\frac{D-4}{r}+\frac{f'}{2f})\partial_r C
\\
\nonumber
&&
+\frac{1}{2}(\frac{h'}{h}-\frac{f'}{f}-\frac{2(D-4)}{r})\partial_rG
+\frac{(D-2)(D-4)}{2r^2f}e^{2B-2C}
-\frac{(D-4)h}{2r^2f}e^{2B-4C+2G}
\\
\nonumber
&&
-\frac{ (D-3)f'}{2rf}
-\frac{f'h'}{4fh}
+\frac{h'}{2rh}
+\frac{h'^2}{4h^2}
-\frac{(D-3)(D-4)}{2r^2}
=0,
\end{eqnarray}
\begin{eqnarray}
\label{ec3}
&&
\hat O^2 C
+(D-4)(\hat O C)^2
+\hat O C \cdot \hat O G
+\hat O A \cdot \hat O C
+\frac{1}{r}(\partial_rA+\partial_r G)
\\
\nonumber
&&
+(\frac{2D-7}{r}+\frac{f'}{f})\partial_rC
-\frac{(D-2)}{r^2f}e^{2B-2C}
+\frac{2h}{r^2f}e^{2B-4C+2G}
+\frac{f'}{rf}
+\frac{D-4}{r^2}=0,
\end{eqnarray}
\begin{eqnarray}
\label{ec4}
 &&
\hat O^2 G
+ (\hat O G)^2
+\hat O A \cdot \hat O G
+(D-4)\hat O C \cdot \hat O G
+\frac{e^{-2A+2G}r^2 h^2}{2f}(\hat O W)^2
\\
\nonumber
&&
+\frac{(D-4)}{2}(\frac{2}{r}+\frac{h'}{h}) \partial_r C
+(\frac{f'}{f}+\frac{h'}{2h}+\frac{(D-2)}{r} )\partial_r G
+(\frac{1}{r}+\frac{h'}{2h})\partial_r A
-\frac{(D-4)h}{r^2f}e^{2B-4C+2G}
\\
\nonumber
&&
+\frac{f'}{rf}
+(\frac{D-3}{2r}+\frac{f'}{2f}-\frac{h'}{2h})\frac{h'}{h}
+\frac{h''}{2h}
+\frac{D-4}{r^2}=0,
\end{eqnarray}
\begin{eqnarray}
\label{ec5}
 \hat O^2 W
-\hat O A \cdot \hat O W
+(D-4)\hat O C \cdot \hat O W
+3\hat O G \cdot \hat O W
+(\frac{D-1}{r}+\frac{2h'}{h})\partial_r W=0~,
\end{eqnarray}
where we define 
\begin{eqnarray}
\label{rel}
\hat O U \cdot \hat O V=\partial_r U \partial_r V+\frac{1}{f}\partial_z U \partial_z V,~~~
\hat O^2 U=\partial_r^2U+\frac{1}{f}\partial_z^2 U,
\end{eqnarray}
and a prime denotes the derivative with respect to $r$.

All other Einstein equations except for  $G_z^r=0$ and $G_r^r-G_z^z=0$
are linear combinations of those used to derive
the above equations or are identically zero.
The remaining Einstein equations $G_z^r=0,~G_r^r-G_z^z=0$
yield two constraints. Following \cite{Wiseman:2002zc}, we note that
setting $G^t_t=G^{\theta_i}_{\theta_i}=G^{\vphi_k}_{\vphi_k}=G^r_r+G^z_z=0$
in $\nabla_\mu G^{\mu r}=0$ and $\nabla_\mu G^{\mu z}=0$, we obtain
\begin{eqnarray}
\partial_z\left(\sqrt{-g} G^r_z \right) +
\sqrt{f} \partial_r\left( \sqrt{f}\sqrt{-g} \frac{1}{2}(G^r_r-G^z_z) \right)
& = & 0 ,
\\
\nonumber 
\sqrt{f}\partial_r\left(\sqrt{-g} G^r_z \right)
-\partial_z\left( \sqrt{f}\sqrt{-g} \frac{1}{2}(G^r_r-G^z_z) \right)
& = & 0 ,
\end{eqnarray}
and, defining $\hat{r}$ via
$\partial/\partial_{\hat{r}} = \sqrt{f}\partial/\partial_{r}$,
then yields the Cauchy-Riemann relations
\begin{eqnarray}
\partial_z\left(\sqrt{-g} G^r_z \right) +
\partial_{\hat{r}}\left( \sqrt{f}\sqrt{-g} \frac{1}{2}(G^r_r-G^z_z) \right)
& = & 0 ,\\
\nonumber 
\partial_{\hat{r}}\left(\sqrt{-g} G^r_z \right)
-\partial_z\left( \sqrt{f}\sqrt{-g} \frac{1}{2}(G^r_r-G^z_z) \right)
& = & 0  .
\end{eqnarray}
Thus the weighted constraints satisfy Laplace equations,
and the constraints are fulfilled,
when one of them is satisfied on the boundary 
and the other at a single point
\cite{Wiseman:2002zc}.

\subsection{General properties }
We impose the event horizon to reside at a surface of constant radial coordinate
$r=r_0$, where $f(r)=f'(r_0)(r-r_0)+O(r-r_0)^2$ while $h(r_0)>0,~f'(r_0)>0$.
Also, the functions $f(r)$ and $h(r) $ take only positive 
values for $r>r_0$ and tend to one for $r \to \infty$.

The Killing vector  
$\chi=\partial/\partial_t+ \sum_k \Omega_k \partial/\partial \varphi_k $
is orthogonal to and null on the horizon. 
For the solutions within the ansatz (\ref{metric}),
the event horizon angular velocities are equal, $\Omega_k=\Omega_H=W(r,z)|_{r=r_0}$.


Utilizing the reflection symmetry of the nonuniform black strings 
w.r.t.~$z=L/2$,
the nonuniform solutions are constructed subject to the following set of 
boundary conditions
\begin{eqnarray}
\label{bc1} 
A\big|_{ r=\infty}=B\big|_{ r=\infty}=C\big|_{ r=\infty}
=G\big|_{ r=\infty}=W\big|_{ r=\infty}=0,
\end{eqnarray}
\begin{eqnarray}
\label{bc2} 
B\big|_{r={r_0}}-A\big|_{r={r_0}}=d_0,~\partial_{ r} 
A\big|_{r={r_0}}=\partial_{r} C\big|_{r={r_0}}
=\partial_{r} G\big|_{r={r_0}}=0,~~
W\big|_{r={r_0}}=\Omega_H, 
\end{eqnarray}
\begin{eqnarray}
\label{bc3} 
\partial_z A\big|_{z=0,L/2}=\partial_z B\big|_{z=0,L/2}=\partial_z C\big|_{z=0,L/2}
=\partial_z G\big|_{z=0,L/2}=\partial_z W\big|_{z=0,L/2}=0, 
\end{eqnarray}
where the constant $d_0$ is related to the Hawking 
temperature of the solutions. 

As in the case of $3+1$-dimensional Kerr black holes,
the rotating black strings have an ergosurface
inside of which
observers cannot remain stationary, and will 
move in the direction of the rotation. 
The ergosurface is located at $g_{tt}=0$, i.e.
\begin{eqnarray} 
\label{er}  
e^{2G(r,z)} r^2 h(r) W(r,z)^2-\frac{e^{2A(r,z)} f(r)}{h(r)}=0 ,
\end{eqnarray} 
and does not intersect the horizon.
(Note that the ergoregion here extends nontrivially in the extra dimension.)

The computation of the conserved charges of rotating black strings
was discussed e.g.~in \cite{Townsend:2001rg}.
The essential idea there is to consider the asymptotic
values of the gravitational field far away from the black string
and to compare them with those corresponding to a gravitational field
in the absence of the black string.
The obvious choice of the background in this case is ${\cal M}^{D-1}\times S^1$, 
the asymptotic form of the relevant metric components 
being
\begin{eqnarray}
\label{1} 
g_{tt}\simeq -1+\frac{c_t}{r^{D-4}},~~~g_{zz}\simeq 1+\frac{c_z}{r^{D-4}},~~~
g_{\varphi_k t}\simeq \left( \prod_{l=0}^{k-1} \cos^2 \theta_l
  \right) \sin^2\theta_k \frac{c_\varphi}{r^{D-4}},
\end{eqnarray}
which reveals the existence of three free parameters $c_t,~c_z$ and $c_\varphi$.
The mass-energy $E$, the tension
${\mathcal T}$ and the angular momenta $J_k$ 
of black string solutions are given by
\begin{eqnarray}
\label{gen-def} 
\hspace{-0.5cm}
E=\frac{A_{D-3}L}{16 \pi G_D}((D-3)c_t-c_z),
~{\mathcal T}=\frac{A_{D-3}}{16 \pi G_D}(c_t-(D-3)c_z),
~J_k=J=-\frac{A_{D-3}L}{8 \pi G_D} c_\varphi, 
\end{eqnarray}
where 
$A_{D-3}=2\pi ^{\frac{D-2}{2}}/\Gamma ({(D-2)}/2)$ is 
the area of the unit $D-3$ sphere.

The global charges of a black string can also 
be computed by using the quasilocal
tensor of Brown and York \cite{Brown:1992br}, 
augmented by the counterterms formalism.
In this approach we add to the action (\ref{action-grav})
  suitable counterterms $I_{ct}$
built up with
curvature invariants of  the induced metric on the boundary $\partial \cal{M}$
\cite{Kraus:1999di},\cite{Astefanesei:2006zd},\cite{Mann:2005yr}.
These counterterms  do not alter the bulk equations of motion.
Unlike the background substraction, this procedure is 
satisfying since it is intrinsic to the spacetime of interest  and it is 
unambiguous once the counterterm is specified.
Our choice of the counterterm was similar to the static case \cite{Kleihaus:2006ee},
$I_{ct}=
-\frac{1}{8 \pi G_D}\sqrt{(D-3)/{(D-4)} }\int_{\partial\mathcal{M}} d^{D-1} x\sqrt{-h}
\sqrt{ \mathcal{R}},$
where  
$\mathcal{R}$ is the Ricci scalar of the boundary geometry.
The variation of the total action with respect to the
boundary metric $h_{ij}$ provides a boundary stress-tensor, whose 
expression is given e.g. in \cite{Kleihaus:2006ee}.
The mass-energy, tension and angular momenta are the charges associated to $\partial/\partial t$,
$\partial/\partial z$ and 
$\partial/\partial \varphi_k$ respectively (note that $\partial/\partial z$
is a Killing symmetry of the boundary metric).
We have verified that for $D=6,~8$ and $D=10$ the expressions computed in this way
agree with (\ref{gen-def})
(see \cite{Astefanesei:2006zd,Astefanesei:2005ad} for similar computations
for a different type of rotating solutions).

One can also define a relative tension $n$ 
(also called the relative binding energy or scalar charge) 
\begin{eqnarray}
\label{3}
n=\frac{{\mathcal T} L}{E}=\frac{c_t-(D-3)c_z}{(D-3)c_t-c_z},
\end{eqnarray}
which measures how large the tension is relative to the mass-energy, being constant
for UBS solutions.

The Hawking temperature $T_H=\kappa_H/2\pi$ can be obtained from
the standard relation
\begin{eqnarray} 
\label{kappa} 
\kappa_H^2=- \left. \frac{1}{2}\nabla^a \chi^b \, \nabla_a\chi_b
\right|_{r=r_0},
\end{eqnarray}
where $\kappa_H$ is the surface gravity, which is constant
at the horizon.
One finds 
\begin{eqnarray}
\label{temp} 
T_H=e^{A_0-B_0}\,T_{H}^{(0)} = e^{-d_0}\,T_{H}^{(0)}, 
\end{eqnarray}
where $T_{H}^{(0)} $ is the Hawking temperature of the uniform solution
based on the same `background' functions (\ref{MP}),
$T_{H}^{(0)}=f'(r_0)/(4\pi  {\sqrt{h(r_0)}})$.
Here and below
 $A_0(z),B_0(z),C_0(z),G_0(z)$ and $W_0(z)$  denote the values of 
the metric functions on the event horizon $r=r_0$.

The area $A_H$ of the black string horizon can also be expressed in a similar way
\begin{eqnarray}
\label{A} 
A_H= A_{H}^{(0)}\frac{1}{L}\int_0^L e^{B_0+(D-4)C_0+G_0}dz,
\end{eqnarray}
with $A_{H}^{(0)}$ the event horizon area of the 
corresponding uniform solution
\begin{eqnarray}
\label{A0} 
A_{H }^{(0)}=LA_{D-3}r_0^{D-3}\sqrt{h(r_0)}~.
\end{eqnarray}
As usual, one identifies the entropy of the black string solutions
with one quarter of their event horizon area, $S=A_H/4G_D$.

Considering the thermodynamics of these solutions,
the black strings should satisfy the first law of thermodynamics 
\begin{eqnarray}
\label{firstlaw}
dE=T_HdS+\frac{(D-2)}{2} \Omega_H dJ+{\mathcal T}dL.
\end{eqnarray}
One may regards the parameters $S,~J$ and $L$ as a complete set of extensive parameters
for the mass-energy $E(S,J,L)$ and define the intensive parameters
conjugate to them.
These quantities are the temperature, the angular velocities and the tension.

Following \cite{Chowdhury:2006qn}, \cite{Kol:2003if}, 
one can derive in a simple way a 
 Smarr formula, by letting the length
of the compact extra dimension change as $L \to L+dL$.
This implies
\begin{eqnarray}
\label{SM1} 
E \to E(1+\frac{dL}{L})^{D-3},~~
S\to S(1+\frac{dL}{L})^{D-2},~~
J\to J(1+\frac{dL}{L})^{D-2}.
\end{eqnarray}
As a result we find the Smarr formula 
(see also \cite{Townsend:2001rg} for a different derivation of this relation)
\begin{eqnarray}
\label{smarrform} 
\frac{D-3-n}{D-2} E=T_HS+\frac{(D-2)}{2} \Omega_H J~.
\end{eqnarray}
 
In the canonical ensemble, we study black strings holding the temperature
$T_H$, the angular momenta $J$
and the length $L$ of the extra dimension fixed.
The associated thermodynamic potential is the Helmholz free energy 
\begin{eqnarray}
\label{F}
F[T_H,J,L]=E-T_HS ~.
\end{eqnarray}
In the grand canonical ensemble, on the other hand, we keep 
the temperature, the angular velocity and the tension fixed.
In this case the thermodynamics is obtained from the Gibbs potential 
\begin{eqnarray}
\label{G}
G[T_H,\Omega_H,{\mathcal T}]=E-T_H S-\frac{(D-2)}{2}\Omega_H J-{\mathcal T} L.
\end{eqnarray}

 We finally remark that
the technique used in \cite{Horowitz:2002dc}, \cite{Harmark:2003eg}
to construct `copies of solutions' works for rotating NUBS, too.
When taking
$f(r)=1-(r_0/r)^{D-4}$, $h(r)=1$, i.e.~(\ref{MP-2}),
 the Einstein equations (\ref{ec1})-(\ref{ec5}) are left invariant
by the transformation $r \to  r/k$, $ z \to z/k$, $ r_0 \to
r_0/k $, with $k$ an arbitrary positive integer.
Therefore, one may generate a family of vacuum solutions
in this way.
The new solutions have the same length of the extra dimension.
Their relevant properties, expressed in terms of the
corresponding properties of the initial solution, read
\begin{eqnarray}
E^{(k)}=\frac{E}{k^{D-4}}, ~T_{H }^{(k)}=k T_H,~~S^{(k)}=\frac{S}{k^{D-3}},~
n^{(k)}=n,~
J^{(k)}=\frac{J}{k^{D-3}},~\Omega_{H }^{(k)}=k\Omega_H.
\end{eqnarray}

\section{Rotating UBS
with equal angular momenta}
\subsection{Thermodynamics}
We start by discussing the properties of 
uniform black string solutions obtained by taking $A=B=C=G=0$ 
in the general ansatz
(\ref{metric}) with the functions $f,h$ and $W(r,z)=w(r)$ given by (\ref{MP}).
The extra dimension plays no role here and the 
general results 
apply to $d=(D-1)$-dimensional MP black holes when formally taking
$L=1$ in the relations below.
 
The uniform black strings
have two parameters $M$ and $a$ which, from (\ref{gen-def}),
are related to the physical mass-energy and angular momenta by
\begin{eqnarray} 
\label{M,a}  
E=\frac{A_{D-3}L}{8\pi G_D}(D-3)M,~~J_i=J=\frac{A_{D-3}L}{4\pi G_D} Ma,
\end{eqnarray} 
Hence one can think of $a$ as essentially  
the angular momentum per unit mass.
The tension ${\mathcal T}$ of the uniform solutions
is fixed by the mass-energy $E$
and length $L$ of the extra dimension 
\begin{eqnarray} 
\label{tens1}  
{\mathcal T}=\frac{E}{L(D-3)}.
\end{eqnarray}

The event horizon of these uniform black strings 
can be determined as the largest root of $1/g_{rr}=0$ 
resp.~$f(r)=0$. That is
\begin{eqnarray} 
\label{eq-f}  
r_0^{D-2}-2Mr_0^2+2Ma^2=0.
\end{eqnarray}
This equation has zero, one or two positive roots, depending
on the sign of $f(r_s)$, where $r_s=(4M/(D-2))^{1/(D-4)}$
is the largest root of  $(r^{D-2}f(r))\,'=0$.
The existence of a regular horizon implies an upper limit on $a$,
\begin{eqnarray} 
\label{f'}  
a^2\leq  \frac{D-4}{D-2} \left(\frac{4M}{D-2}\right)^{\frac{2}{D-4}}
= a^2_{max},
\end{eqnarray}
which via (\ref{M,a}) can also be expressed 
as
\begin{eqnarray} 
\label{bound1}  
 \frac{J}{E^{\frac{D-3}{D-4}}}\leq
 \sqrt{D-4}
\left(\frac{2^{D+1}\pi}{(D-3)^{D-3}(D-2)^{(D-2)/2}}\right)^{1/(D-4)}
(\frac{G_D}{A_{D-3}L})^{1/(D-4)}.
\end{eqnarray}
This strongly contrasts with the case of MP UBS solutions 
with a single nonzero angular momentum,
where for $D>6$ there are configurations with arbitrarily large $J$,
without an occurrence of an extremal limit\footnote{
However, as argued in \cite{Emparan:2003sy}, MP black holes with
a single nonzero angular momentum are in 
fact classically unstable (at least for large rotation) 
and an effective Kerr bound arises through
a dynamical decay mechanism.}.

The ergosurface of the uniform solutions is located at $r_e=(2M)^{1/(D-4)}$.
The horizon angular velocities and the Hawking temperature of the UBS solutions
are given by
\begin{eqnarray} 
\label{omegah}  
\Omega_H=\frac{a}{r_0^2},~~~~T_H=\frac{1}{4\pi r_0}\left(\frac{2M}{r^{D-4}_0}\right)^{1/2}
\left((D-2)\frac{r^{D-4}_0}{2M}-2 \right).
\end{eqnarray}
Solutions with $a=a_{max}$ are extremal black strings with $T_H=0$,
possessing a nonzero entropy.
As can be seen e.g.~by computing the Kretschmann scalar,
the hypersurface $r=r_0$ is not singular in this limit.

 Similar to the static case, the
gravitational thermodynamics of the rotating UBS can be formulated via the Euclidean
path integral.
The Euclidean spinning black string solutions can be obtained
from the Minkowskian ones by sending $t \to -it$
and $a \to ia$ (complexifying $a$ is necessary in order to keep 
the $dt d\varphi_i$ part of the 
metric real).
The thermodynamic system has a constant temperature $T_H=1/\beta$
which is determined by requiring the Euclidean
section be free of conical singularities  (the temperature computed in this way
 coincides with that in (\ref{omegah})).
The  
partition function for the gravitational field is defined by a sum
over all smooth Euclidean geometries which are periodic with a period
$\beta$ in imaginary time \cite{Gibbons:1976ue}.
This integral is computed by using the saddle point approximation,
 the global charges and entropy of the solutions being evaluated 
by standard thermodynamic formulae. 
Upon application of the Gibbs-Duhem relation to the partition 
function \cite{Mann:2003 Found},
this yields an expression for the entropy
\begin{eqnarray}
S=\beta \left(E- \frac{(D-2)}{2}\Omega_H J-{\mathcal T} L\right)-I,  
\label{GibbsDuhem}
\end{eqnarray}%
(with $I$ the regularized tree level Euclidean action),
which agrees with that computed from (\ref{A}).
The entropy can be written in terms of $M$, $r_0$ as
\begin{eqnarray}
\label{entropy} 
S=\frac{1}{4G_D}A_{D-3}Lr_0^{D-3}
\left(\frac{2M}{r^{D-4}_0}\right)^{1/2} .
\end{eqnarray}
The parameters $M$, $a$, $r_0$ can be eliminated and one can write the following 
equation of state (analogous to $f(p,V,T)$, for, say, a gas at pressure $p$ and volume $V$)
\begin{eqnarray} 
\label{n12} 
J=2^{-(D+6)/2}\frac{A_{D-3}L}{G_D}
(D-4)^{D-2}\pi^{1-D}\Omega_H T_H^{2-D}
\left(1+\sqrt{1+\frac{(D-4)\Omega_H^2}{2\pi^2T_H^2}}\right)^{-D/2}
\\
\nonumber
\times
\left((D-2)\sqrt{1+\frac{(D-4)\Omega_H^2}{2\pi^2T_H^2}}-D+6\right)^{-(D-4)/2}.
\end{eqnarray}
The following relations are also useful in what follows
\begin{eqnarray} 
\label{n5} 
T_H=\frac{D-4}{\pi}\left(\frac{A_{D-3}L}{2^{2(D-2)}G_D}\right)^{\frac{1}{D-3}}
S^{1/(3-D)}
\left(1+\frac{4\pi^2 J^2}{ S^2}\right)^{-\frac{D-4}{2(D-3)}}
\left(1-\frac{8 \pi^2 J^2}{(D-4)S^2}\right) ,
\end{eqnarray}
\begin{eqnarray} 
\label{n10} 
E=2^{-\frac{D+6}{2}}(D-3)(D-4)^{D-4}\pi^{3-D}\frac{A_{D-3}L}{G_D}
T_H^{4-D}
\left(1+\sqrt{1+\frac{(D-4)\Omega_H^2}{2\pi^2T_H^2}}\right)^{(2-D)/2}
\\
\nonumber
\times
\left( (D-2)\sqrt{1+\frac{(D-4)\Omega_H^2}{2\pi^2T_H^2}}+6-D\right)^{(6-D)/2}~,
\end{eqnarray}
\begin{eqnarray} 
\label{n11} 
S=2^{-(D+2)/2}\frac{A_{D-3}L}{G_D}
(D-4)^{D-3}\pi^{3-D}T_H^{3-D}
\left(1+\sqrt{1+\frac{(D-4)\Omega_H^2}{2\pi^2T_H^2}}\right)^{-(D-2)/2}
\\
\nonumber
\times
\left((D-2)\sqrt{1+\frac{(D-4)\Omega_H^2}{2\pi^2T_H^2}}-D+6\right)^{-(D-4)/2}~,
\end{eqnarray}
\begin{eqnarray} 
\label{ESJ} 
E= {2^{-\frac{2(D-2)}{(D-3)}}}\frac{(D-3)}{\pi}(\frac{A_{D-3}L}{G_D})^{1/(D-3)}S^{\frac{D-4}{D-3}}
\left(1+\frac{4\pi^2J^2}{S^2}\right)^{\frac{D-2}{2(D-3)}}~.
\end{eqnarray}
As implied by (\ref{n5}), the rotating UBSs have always 
a smaller temperature for the same entropy than
the static UBSs.
The energy to entropy ratio of the UBS solutions satisfies the following bounds
\begin{eqnarray} 
\label{lim1} 
\frac{1}{2^{2(D-2)}}\leq (\frac{\pi}{D-3})^{D-3}
\frac{G_D}{A_{D-3}L}\frac{E^{D-3}}{S^{D-4}}
\leq \frac{(D-2)^{(D-2)/2}}{2^{5(D-2)/2}}~,
\end{eqnarray}
these limits being approached for static and extremal solutions, 
respectively. There is also an upper bound for the ratio $J/S$,
\begin{eqnarray} 
\label{lim2} 
\frac{J}{S}\leq \frac{1}{2\pi} \sqrt{\frac{D-4}{2}}~.
\end{eqnarray}

The analysis of the thermodynamic stability of the rotating UBS
solutions can be performed using the above relations. 
It is known that different thermodynamic ensembles are not exactly
equivalent and may not lead to the same conclusions
as they correspond to different
physical situations. 
Mathematically, thermodynamic stability is equated with the subadditivity 
of the entropy function.
This requires $S(E,J,L)$ to be a concave function of its extensive variables.
The stability can also be studied by the behaviour of the energy $E(S,J,L)$
which should be a convex function.
Therefore one has to compute the determinant of the Hessian matrix of
$E(S,J,L)$ with respect to its extensive variables $X_i$, 
$H^{E}_{X_iX_j}=[\partial^2E/\partial X_i\partial X_j]$ 
\cite{Cvetic:1999ne}, \cite{Caldarelli:1999xj}.
 
In the canonical ensemble, the subadditivity of the entropy
is exactly equivalent to positivity
of the specific heat at constant $(J,L)$, $C_{J,L}=T_H(\partial S/\partial T_H)_{J,L}$.
Also, the  Gibbs potential which is relevant for a grand canonical ensemble 
can be written as
 $ 
G[T_H,\Omega_H,L]=E/(D-2),
$
as a result of the Smarr law (\ref{smarrform}).

At this point it is instructive to see first the corresponding situation 
for the $(D-1)$-dimensional  MP black holes.
It is easy to work out from (\ref{n5}) 
 that the condition for a positive specific heat at fixed
angular momenta
 is equivalent to
 \begin{eqnarray}
 \label{lim-JS} 
 32(D-1)\pi^4J^4+4(D(D-5)+10)\pi^2J^2S^2-(D-4)S^4>0.
 \end{eqnarray}
As expected,
 $C_{J}$ is negative far from extremality, becoming positive
for large enough values of $J$.

To discuss the thermodynamic stability of black holes in a grand canonical 
ensemble, we consider first the specific heat at constant 
angular velocity at the horizon 
\begin{equation}
C_\Omega=T_H\left(\frac{\partial S}{\partial T_H}\right)_{\Omega_H}.
\end{equation}
A straightforward computation using (\ref{n11})
shows that this is a negative quantity 
in the full range of variables, $C_\Omega<0$. 
One can also verify that the determinant of the Hessian 
 matrix of
$E(S,J)$ is negative  
\begin{eqnarray}
\label{Hess}
{\rm det} (\partial^2E/\partial X_i\partial X_j)=-
\frac{4(D-2)\pi^2 }{(D-3)^3 }\frac{E^2(8J^2\pi^2+(D-2)S^2)}{S^2(4J^2\pi^2+S^2)}<0~.
\end{eqnarray} 
As a result, all MP rotating black hole solutions with equal angular momenta
are unstable in a grand canonical ensemble,
and also the configurations far from extremality 
in a canonical ensemble.

Another `response function' of interest is the 
`isothermal permittivity'
\begin{eqnarray}
\epsilon_T \equiv \left(\frac{\partial J}{ 
\partial\Omega_H}\right)_{T_H}.
\end{eqnarray}
One finds from (\ref{n12})
that the condition for a positive $\epsilon_T $ is
\begin{eqnarray} 
\label{n13} 
\frac{\Omega_H}{T_H}<  \left(\frac{\pi^2}{D-4}
\left(\frac{2+5D-D^2}{(D-2)(D-3)}
+\sqrt{\frac{D^2-5D+22}{(D-2)(D-3)}}~\right)\right)^{1/2}.
\end{eqnarray}
For $D=6$, this corresponds to $\Omega_H/T_H<1.7165$, for $D=8$ to
$\Omega_H/T_H<0.7892$ and for $D=10$ to $\Omega_H/T_H<0.477$.
All other rotating black hole solutions have a
negative `isothermal permittivity' and thus
are unstable to angular 
fluctuations, both in a grand canonical and a canonical ensemble.

These conclusions remain unchanged when adding one (trivial) 
extra dimension to the black hole solutions.
Similar to the static case, all grand canonical
configurations are thermally unstable\footnote{ However, 
in this case the determinant of the Hessian
vanishes identically, as a result of the special dependence on $L$.
This happens already for a Schwarzschild black string when considered 
as a solution in a grand canonical ensemble with fixed $T_H,~{\mathcal T}$.}.
Also, all rotating UBS solutions  with  
$J/S<J^{(c)}/S^{(c)}$ are unstable  in a canonical ensemble, 
where the critical ratio $ J^{(c)}/S^{(c)}$ is given by
\begin{eqnarray}
\label{t2} 
\frac{J^{(c)}}{S^{(c)}} = \frac{1}{4 \pi \sqrt{D-1}}\left( 
 \sqrt{(D-2)(D-3)(D(D-5)+22)}-D(D-5)-10 \right)^{1/2}.
\end{eqnarray}
At the critical point, the specific heat goes through an infinite discontinuity,
and a second order phase transition takes place.
The critical values for other relevant quantities read
\begin{eqnarray} 
\label{n6} 
T_{H}^{(c)} =f_1(D)(J^{(c)})^{ -\frac{1}{D-3}} ,~~E^{(c)}=f_2(D)(J^{(c)})^{\frac{D-4}{D-3} },
~~\Omega_H^{(c)}=f_3(D)(J^{(c)})^{-\frac{1}{D-3}}~,
\end{eqnarray}
where
\begin{eqnarray} 
\label{n7} 
 f_1(D)&=& 
2^{-\frac{2D-1}{D-3}}(\frac{A_{D-3}L}{G_D})^{1/(D-3)}\pi^{-\frac{D-2}{D-3}}
(D-1)^{-\frac{(D-1)}{2(D-3)}}
\left(f_0(D)-(D-2)(D-7)\right)^{-\frac{D-4}{2(D-3)}}
\nonumber
\\
&&\times \left(3(D-2)(D-3)-f_0(D)\right)\left(f_0(D)-18+5D-D^2\right)^{\frac{1}{2(D-3)}},
\nonumber
\\
f_2(D)&=&  2^{-\frac{D+2}{D-3}}(\frac{A_{D-3}L}{G_D})^{1/(D-3)}
\pi^{1/(3-D)}(D-3)(D-1)^{1/(3-D)}
\\
\nonumber
&&\times(f_0(D)-(D-2)(D-7))
^{\frac{D-2}{2(D-3)}}(f_0(D)-D(D-5)-10)^{-\frac{D-4}{2(D-3)}}~,
\\
\nonumber
f_3(D)&=& (32\pi)^{1/(3-D)}(\frac{A_{D-3}L}{G_D})^{1/(D-3)}(D-1)^{1/(3-D)}
(f_0-(D-2)(D-7))^{-\frac{D-4}{2(D-3)}}
\\
\nonumber
&&\times
(f_0(D)-D(D-5)-10)^{\frac{D-2}{2(D-3)}}~,
\end{eqnarray}
while
\begin{eqnarray} 
\label{n8-s} 
f_0(D)=\sqrt{(D-2)(D-3)(D(D-5)+22)}~.
\end{eqnarray}
One finds for example the critical ratio 
$\Omega_H^{(c)}/T_{H}^{(c)}\simeq 2.42757$ for $D=6$,
$\Omega_H^{(c)}/T_{H}^{(c)}\simeq 1.1162$ for $D=8$ and 
$\Omega_H^{(c)}/T_{H}^{(c)}\simeq 0.6747$ for $D=10$.

\subsection{Gregory-Laflamme instability }

It is natural to expect that the MP uniform black strings become
unstable at critical values of the mass and angular momentum.
To determine these critical values for 
black strings with equal magnitude angular momenta,
we make an expansion around the UBS of the form
\begin{eqnarray}
\label{p-ans} 
\nonumber
&&A(r,z)= \epsilon a_1(r) \cos(kz) +O(\epsilon^2) ,
\\
\nonumber
&&B(r,z)=\epsilon b_1(r) \cos(kz) +O(\epsilon^2),
\\
\nonumber
&&C(r,z)=\epsilon c_1(r) \cos(kz)+O(\epsilon^2) ,
\\
\nonumber 
&&G(r,z)=\epsilon g_1(r) \cos(kz)+O(\epsilon^2) ,
\\
&& W(r,z)=w(r)+\epsilon w_1(r) \cos(kz)+O(\epsilon^2),
\end{eqnarray}
\begin{figure}[t!]
\setlength{\unitlength}{1cm}
\begin{picture}(8,8)
\put(1.5,0.0){\epsfig{file=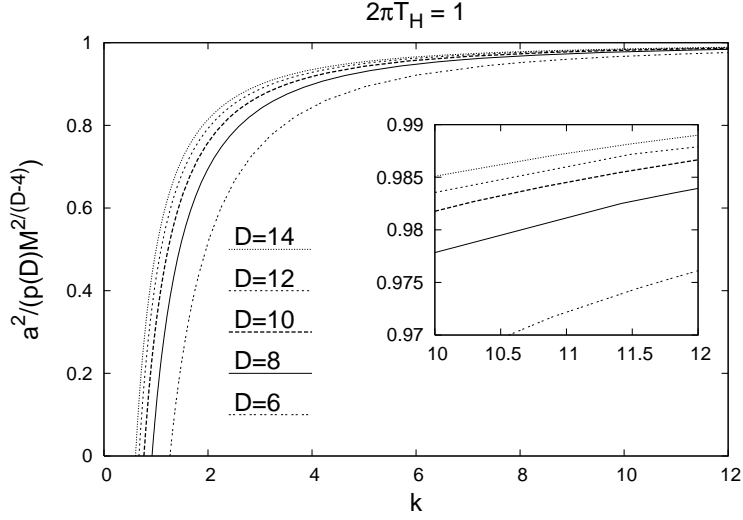,width=10cm}}
\end{picture}
\caption{
The dimensionless quantity $a^2/ \left( p(D) M^{\frac{2}{D-4}} \right)$
is shown as a measure of the rotation of the uniform black strings 
at constant temperature $2 \pi T_{H}=1$
in $D$ even dimensions, $6 \le D \le 14$,
versus the wavenumber $k$ of the zeromode fluctuation.
}
\end{figure}
with $\epsilon$ a small parameter and $f,h,w$ given by (\ref{MP}).
%
This expansion is appropriate for studying perturbations at the wavelength 
which is marginally stable.

Upon substituting the above expressions   into the Einstein equations,
to order $O(\epsilon)$
the following set of ODE is generated 
\begin{eqnarray}
\label{p1} 
\nonumber
& 
a_1''
+(\frac{D-3}{r}+\frac{3f'}{2f}-\frac{h'}{2h})a_1'
+( \frac{f'}{2f}-\frac{h'}{2h})((D-4)c_1'+g_1')
-\frac{r^2h^2w'}{ f}w_1'
+\frac{r^2 h^2w'^2}{f}(a_1-g_1)-\frac{k^2a_1}{f}=0,
\\
\nonumber
& 
b_1''
-(\frac{D-3}{r}+\frac{h'}{2h})a_1'
-\frac{(D-4)^2}{r}c_1'
+\frac{f'}{2f}(b_1'-(D-4)c_1')
+\frac{1}{2}(\frac{h'}{h}-\frac{f'}{f}-\frac{2(D-4)}{r})g_1'
\\
\nonumber
&
-\frac{r^2h^2w'}{2f}w_1'
-\frac{(D-4)h}{r^2f}(b_1-2c_1+g_1)
+\frac{r^2h^2w'^2}{2 f}(a_1-g_1)
+\frac{(D-2)(D-4)}{r^2f}(b_1-c_1)
-\frac{k^2b_1}{f}=0,
\\
\nonumber
&c_1'' 
+\frac{f'}{f}c_1'
+\frac{1}{r}(a_1'+(2D-7)c_1'+g_1')
+\frac{2(D-2)}{r^2f}(c_1-b_1)
+\frac{4h}{r^2f}(b_1-2c_1+g_1)
-\frac{k^2c_1}{f}=0,
\\
\label{p4} 
&
g_1'' 
+ (\frac{D-2}{r}+\frac{f'}{f}+\frac{h'}{2h})g_1'
+\frac{r^2h^2w'}{f}w_1'
+\frac{2(D-4)h}{r^2f}(2c_1-g_1-b_1)
\\
\nonumber
&
+(\frac{1}{r}+\frac{h'}{2h})(a_1'+(D-4)c_1')
+\frac{r^2 h^2w'^2}{f}(g_1-a_1)-\frac{k^2g_1}{f}=0,
\\
\nonumber
&
w_1''  
+(\frac{D-1}{r}+\frac{2h'}{h})w_1'
+(-a_1'+(D-4)c_1'+3g_1')w'
-\frac{k^2w_1}{f}=0.
\end{eqnarray}
This eigenvalue problem for the wavenumber $k=2\pi/L$ is then solved numerically
with suitable boundary conditions, 
for rotating black strings in $D$ even dimensions, $6 \le D \le 14$.
The results are displayed in Figure 1,
where we exhibit the dimensionless quantity (see Eqs.~(\ref{M,a}), (\ref{f'}))
$a^2/\left( p(D) M^{\frac{2}{D-4}} \right)$, with
 \begin{eqnarray} 
 \label{fig'}  
 p(D)  =
 \frac{D-4}{D-2} \left(\frac{4 }{D-2}\right)^{\frac{2}{D-4}}, 
 \end{eqnarray}
as a measure of the rotation
versus the wavenumber $k$ for constant temperature,
$2 \pi T_{H}=1$.
Note, that for extremal solutions 
$a^2/\left( p(D) M^{\frac{2}{D-4}} \right)=1$.
Keeping the temperature fixed, the wavenumber $k$ of the marginally
stable mode increases with increasing rotation
and decreases with increasing dimension $D$.

Introducing the scaled energy-mass $M_s$ and the scaled
angular momentum $J_s$ (following Eq.~(\ref{gen-def}))
\begin{eqnarray}
\label{gen-defsc}
\hspace{-0.5cm}
E=
 \frac{A_{D-3}L}{16 \pi G_D}(D-3) M_s,
~J_k=J=
-\frac{A_{D-3}L}{8 \pi G_D} J_s,
\end{eqnarray}
these results are presented in Figure 2 (left) in a ``phase diagram'' format 
for fixed $L=L_0$, where $L_0$ there corresponds to the critical length of the corresponding
static solutions.
Here $M_s$ and $J_s$ are suitably normalized and equipped with powers, such that
the extremality curve is the same for any dimension $D$.
 
In principle, following \cite{Gregory:1993vy},
one can get an estimation of $k$ by equating 
the entropy of the rotating string
with that of a MP rotating black hole with the same momenta.
The entropy, mass-energy and angular momenta of a black string
are related through (\ref{ESJ}).
The corresponding relation 
for a MP black hole in $D$ dimensions with $(D-2)/2$ equal angular
momenta  can easily be derived by using the
relations in \cite{Myers:1986un}, and reads
\begin{eqnarray}
\label{MSJ-BH}
E=\frac{(D-2)}{\pi}2^{-\frac{2(D-1)}{D-2}}\left(\frac{A_{D-2}}{G_D}\right)^{1/(D-2)}
S^{\frac{D-3}{D-2}}\sqrt{1+\frac{4\pi^2 J^2}{S^2}}~.
\end{eqnarray}
\begin{figure}[t!]
\setlength{\unitlength}{1cm}
\begin{picture}(8,6) 
\put(-0.9,0.0){\epsfig{file=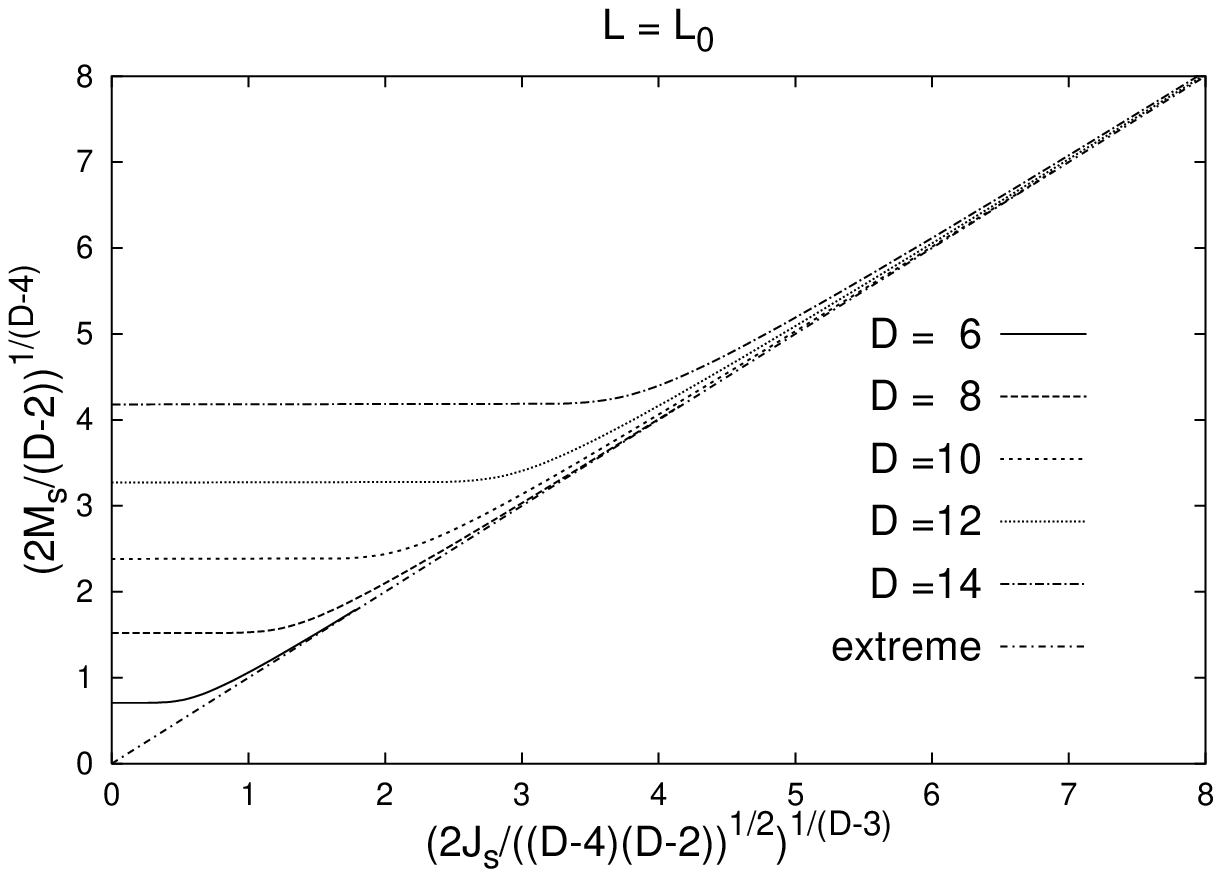,width=8.cm}}
\put(7.7,0.0){\epsfig{file=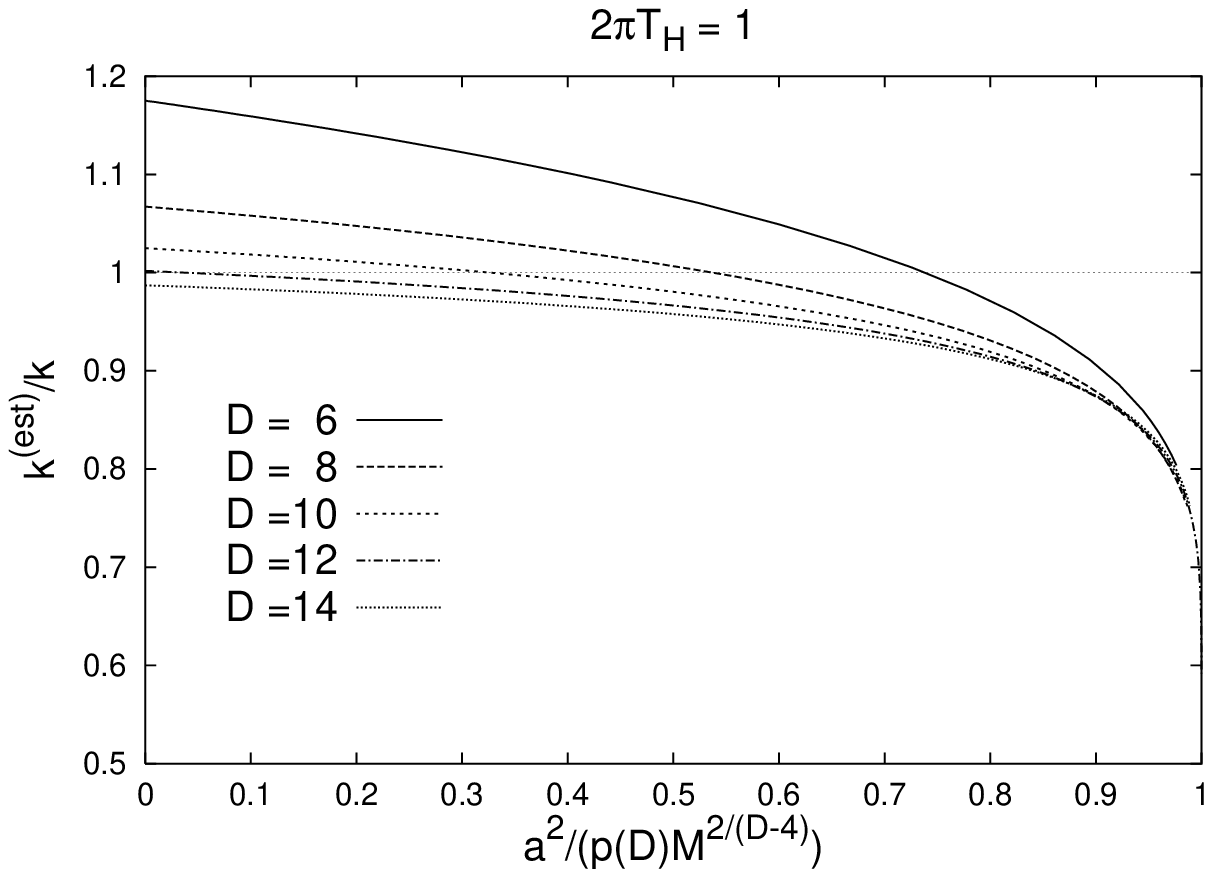,width=8.4cm}}
\end{picture}
\caption{
Left: 
The relation between the scaled mass  $M_s$ and scaled angular momentum $J_s$ 
for fixed critical length of the extra dimension $L$
 is shown for rotating solutions
in $D$ even dimensions, $6 \le D \le 14$.
Both $M_s$ and $J_s$ are 
equipped with suitable powers and normalizations.
$L_0$ represents the value
where the instability of the static uniform black string occurs.
Right: The ratio betwen the wavelength  estimate $k^{(est)}$
and the value of $k$ found numerically 
is plotted at constant temperature $2 \pi T_{H}=1$
as function
of the dimensionless quantity $a^2/ \left( p(D) M^{\frac{2}{D-4}} \right)$.
}
\end{figure}
The expression for the wavelength estimate  $k^{(est)}$
we find 
  by equating 
the entropies is
\begin{eqnarray}
\label{k-est}
k^{(est)}=  k^{(est)}_{st}
\left(1+\frac{4\pi^2 J^2}{S^2}\right)^{\frac{D-2}{2(D-3)}} 
\end{eqnarray}
where 
\begin{eqnarray}
\label{k-est0}
k^{(est)}_{st}= 2^{-\frac{D+1}{ D-3 }}\pi^{-\frac{D-2}{D-3}}
\frac{ A_{D-3}}{(A_{D-2})^{\frac{ D-4 }{D-3}}}
\frac{(D-3)^{D-3}}{(D-2)^{\frac{(D-4)(D-2)}{D-3}}}
\frac{1}{(G_DE)^{1/(D-3)}}
\end{eqnarray}
is the wavelength estimate
for a static solution with the same value of $E$.
From (\ref{lim2}) we find that nonextremal rotating solutions satisfy the inequality
\begin{eqnarray}
 k^{(est)}_{st}\leq k^{(est)} < 
\left(\frac{D-2}{2}\right)^{\frac{D-2}{2(D-3)}}k^{(est)}_{st}~,
\end{eqnarray}
and thus $k^{(est)}$ stays finite in the extremal limit.

The above wavelength estimate   can also be expressed in 
terms of variables used in the numerical
procedure as
\begin{eqnarray}
\label{k-estn}
k^{(est)}= 2 \pi \left(\frac{D-3}{D-2}\right)^{D-2}
\frac{A_{D-3}}{A_{D-2}}
\frac{1}{\sqrt{r_0^2-a^2}}~.
\end{eqnarray}
The fact that, as seen in Figure 1, the numerical
value of $k$  takes very large values
in the limit 
$a^2/\left( p(D) M^{\frac{2}{D-4}} \right)\to 1$,
 is 
a consequence of keeping the temperature constant and letting $r_0$ run.
If we would rescale with $r_0$, the temperature would vanish in the extreme limit 
and the rescaled $k$ would stay finite.

The ratio between the numerical value of $k$ one finds by solving
the equations (\ref{p4}) and the above estimate is presented in 
 Figure 2 (right).
 
For static black strings, the study of the perturbative equations 
in second order revealed the appearance of a critical dimension,
above which the perturbative nonuniform black strings
are less massive than the marginally stable uniform black string
\cite{Sorkin:2004qq}.
It would therefore be interesting to solve the perturbative equations
to second order also in the presence of rotation. 
But so far we have encountered numerical problems in such an analysis.

\section{
\boldmath $\hspace{-0.1cm} D=6$ \unboldmath
rotating nonuniform black string solutions}

\subsection{Numerical procedure}
 
To construct rotating nonuniform black string solutions numerically,
we introduce analogous to the static case
the new radial coordinate $\tilde r$, 
\begin{equation}
\tilde r=\sqrt{r^2 - r_0^2 } \ ,
\end{equation}
thus the horizon resides at $\tilde r=0$.

The $D=6$ line element Eq.~(\ref{metric}) then reads
(with $\theta_1=\theta$)
\begin{eqnarray}
\label{metric6D}
 ds^2&=&
-\,  e^{2 A(\tilde r,z)} \, \frac{\tilde r^2}{g(\tilde r)}\, dt^2
+ e^{2 B(\tilde r,z)}\, (d\tilde r^2+dz^2)
+ e^{2 C(\tilde r,z)}\, g(\tilde r)\, d\theta^2
\\
\nonumber
&&
+\, e^{2 G(\tilde r,z)}\, g(\tilde r)
 \bigg(\sin^2 \theta \, (d \varphi_1- W(\tilde r,z)\, dt)^2 
      +\cos^2 \theta \, (d \varphi_2- W(\tilde r,z)\, dt)^2 
 \bigg) 
\\
\nonumber
&&
-\, (e^{2 G(\tilde r,z)}-e^{2 C(\tilde r,z)})
 \, g(\tilde r)\sin^2 \theta \cos^2 \theta
\, (d \varphi_1-d \varphi_2)^2 \ ,
\end{eqnarray}
where
$g(\tilde r)=r_0^2+\tilde r^2$.

We then change to dimensionless coordinates
$\rho$ and $\zeta$,
\begin{eqnarray}
\rho = \tilde r/(r_0+\tilde r)  , \ \ \ \zeta = z/ L  ,
  \label{barx2} \end{eqnarray}
where the compactified radial coordinate $\rho$
maps spatial infinity to the finite value $\rho=1$,
and $L$ is the asymptotic length of the compact direction.

We solve the resulting set of five coupled non-linear
elliptic partial differential equations numerically,
subject to the boundary conditions Eqs.~(\ref{bc1})-(\ref{bc3}).
These numerical calculations are based on the Newton-Raphson method
and are performed with help of the program FIDISOL \cite{schoen},
which provides also an error estimate for each unknown function.

The equations are discretized on a non-equidistant grid in
$\rho$ and $\zeta$.
Typical grids used have sizes $65 \times 50$,
covering the integration region
$0\leq \rho \leq 1$ and $0\leq \zeta \leq 1/2$.
(See \cite{schoen} and \cite{kk}
for further details and examples for the numerical procedure.)
For the nonuniform strings the estimated relative errors 
range from approximately $\approx 0.001$\%
for small geometric deformation
to $\approx 1$\% for large deformation. 
We also monitored the violation of the weighted constraints
 $    \sqrt{f} \sqrt{-g} (G_r^r-G_z^z)=0, $ and
     $  \sqrt{-g} G_z^r=0 \ ,
 $
which is typically less then $0.1$.

The horizon coordinate $r_0$ and the asymptotic length $L$ of the compact
direction enter the equations of motion as parameters.
The results presented are obtained with the parameter choice
\begin{equation}
r_0=1 \ , \ \ \ L=L^{\rm crit}= 
4.9516 
, \label{r_0-L} \end{equation}
where $L^{\rm crit}$ represents the value, 
where the instability of the static uniform black string occurs.

Rotating nonuniform black strings can then be obtained by
starting from a static nonuniform black string solution
and increasing the value of angular velocity $\Omega_H$
of the event horizon, which enters the boundary conditions.
By varying also the second boundary parameter $d_0$,
associated with the temperature of the black strings,
$d_0=\ln (T_{H}^{(0)}/T_H)$ (see Eq.~(\ref{temp})),
the full set of rotating nonuniform black strings
can then be explored.
An alternative procedure to obtain rotating NUBS 
numerically would be, to start from stationary 
perturbative nonuniform solutions.

The basic properties of the NUBS are encoded in
the five metric functions $A(\rho,\zeta)$, $B(\rho,\zeta)$,
$C(\rho,\zeta)$, $G(\rho,\zeta)$, and $W(\rho,\zeta)$.
These functions
change smoothly with the two boundary parameters $d_0$ and $\Omega_H$.
We illustrate these functions in Figure 3, 
for the parameter choices $d_0=0.6$ and $\Omega_H=0.25$ 
resp.~$\Omega_H=0.202$, the latter
corresponding to a solution in the strongly deformed region
close to the expected transition 
from rotating nonuniform black strings
to rotating caged black holes.

\begin{figure}[t!]
\setlength{\unitlength}{1cm}
\begin{picture}(15,18)
\put(-1,0){\epsfig{file=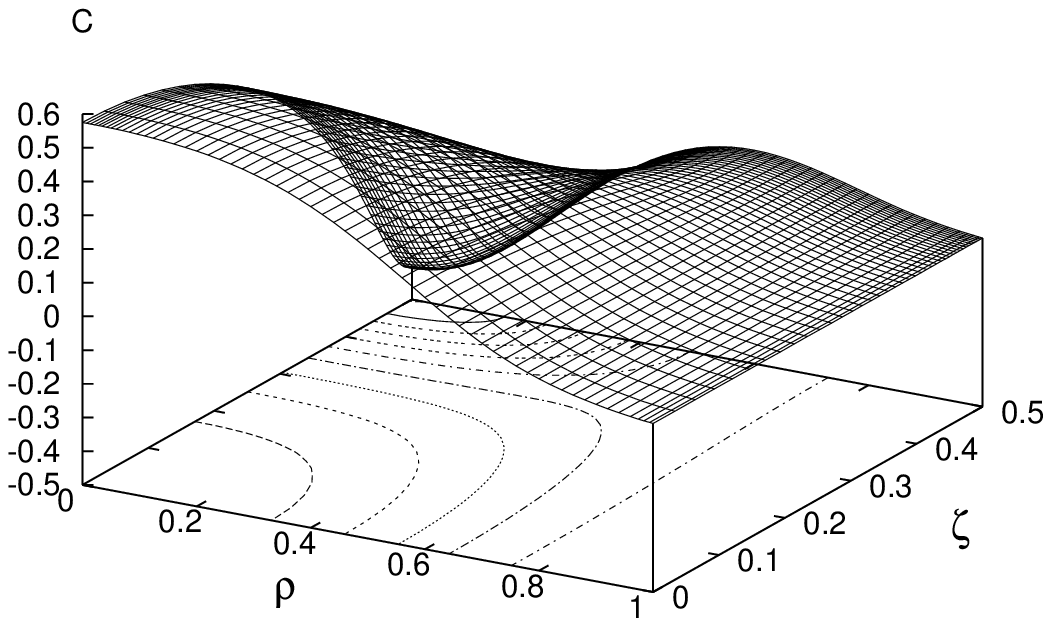,width=8cm}}
\put(7,0){\epsfig{file=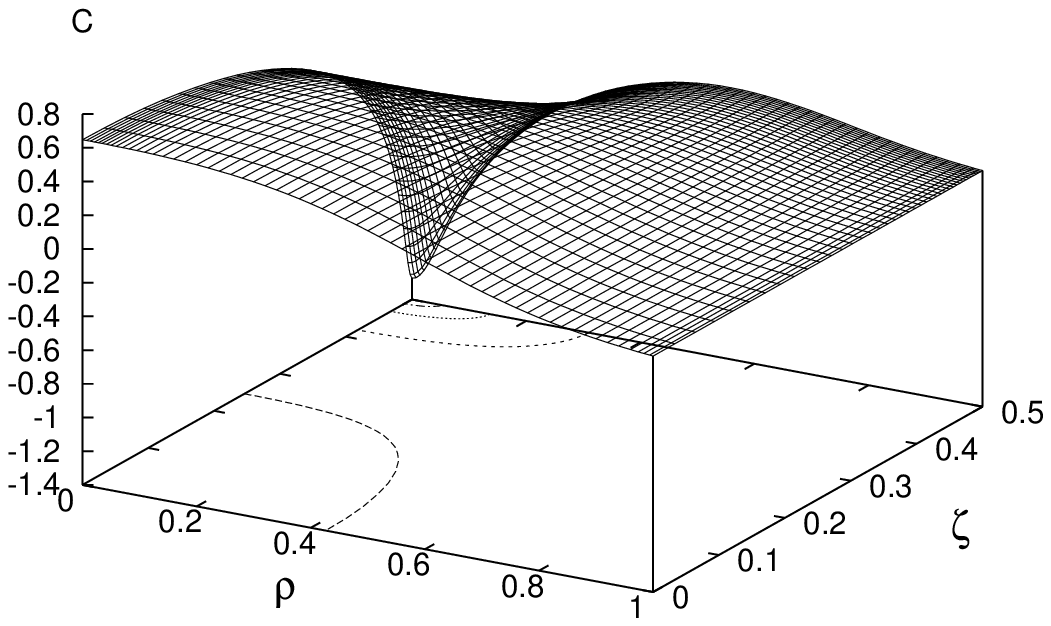,width=8cm}}
\put(-1,6){\epsfig{file=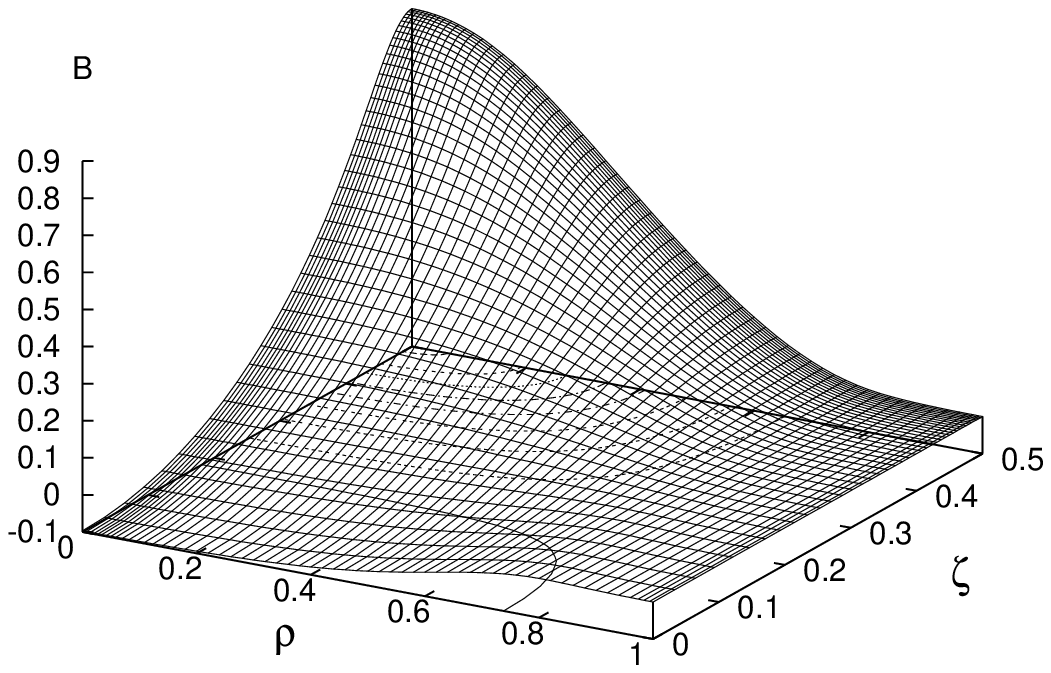,width=8cm}}
\put(7,6){\epsfig{file=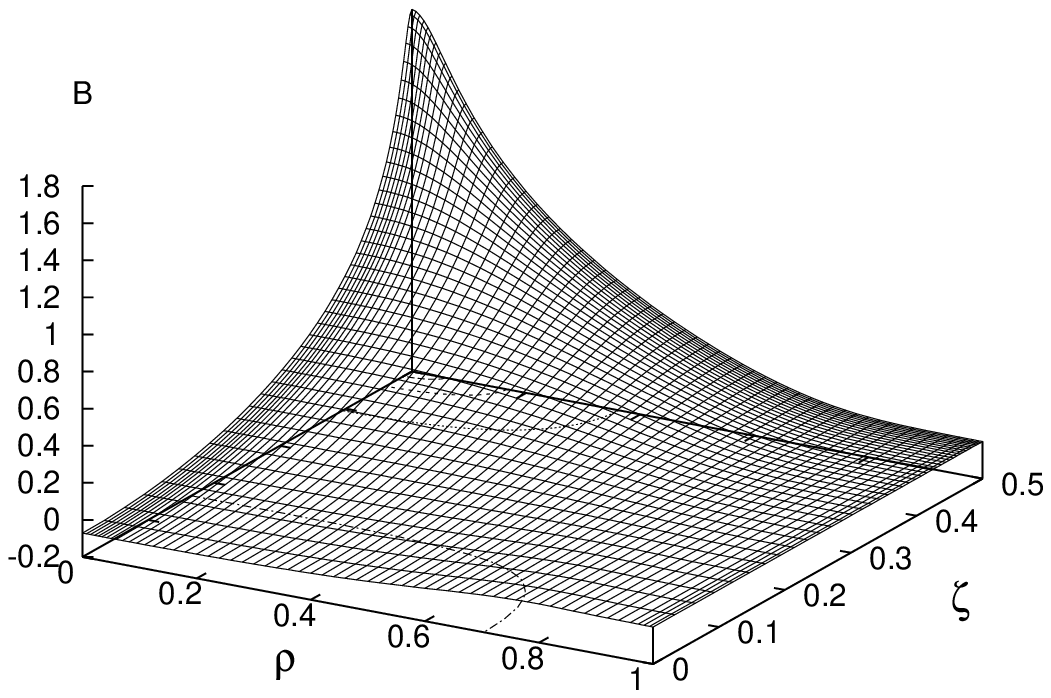,width=8cm}}
\put(-1,12){\epsfig{file=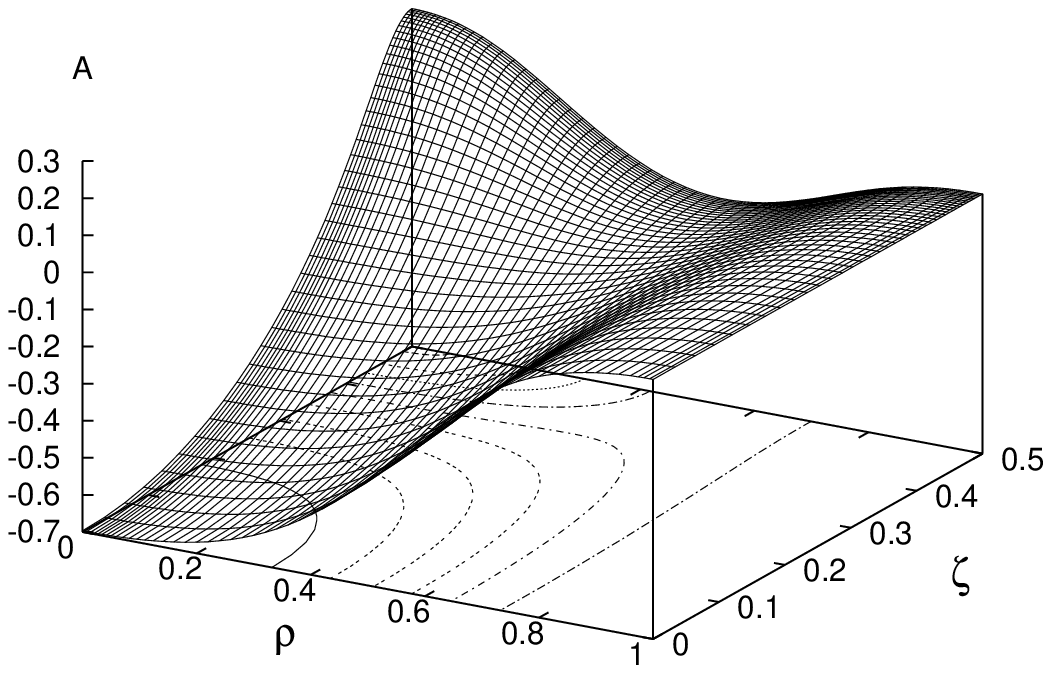,width=8cm}}
\put(7,12){\epsfig{file=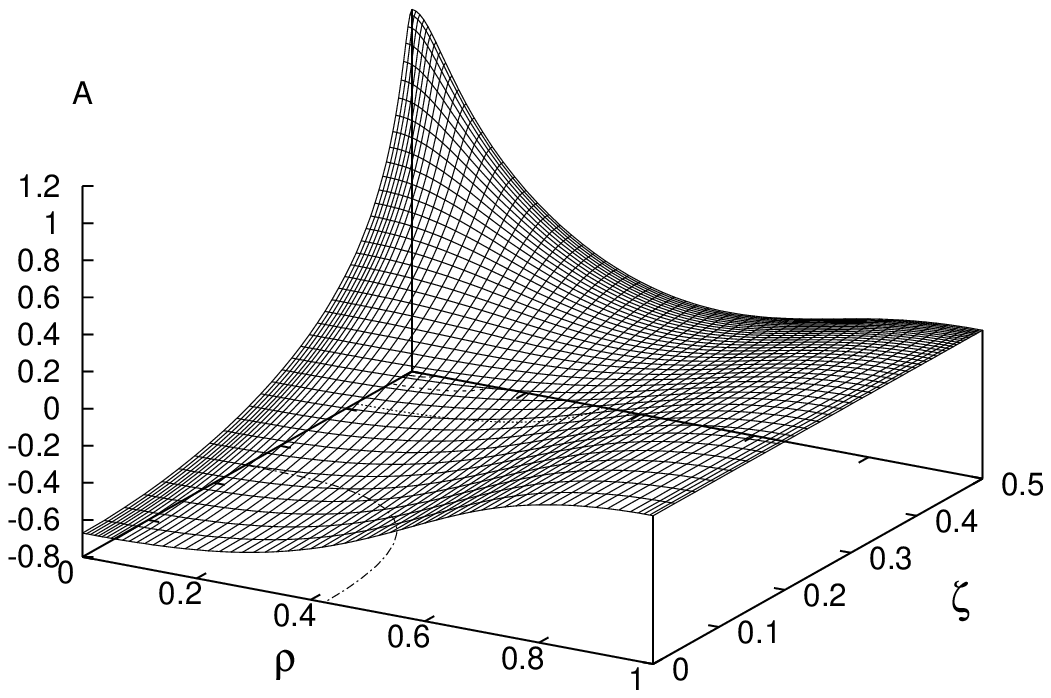,width=8cm}}
\end{picture}
\caption{
The metric functions $A$, $B$, $C$, $G$ and $W$
of the $D=6$ rotating nonuniform black string solution
with temperature parameter $d_0=0.6$ and 
horizon angular velocity $\Omega_H=0.25$ (left column)
and $\Omega_H=0.202$ (right column)
are shown as functions of the compactified radial coordinate $\rho$,
and the coordinate $\zeta$ of the compact direction.
Note that the horizon is located at $\rho=0$.
}
\end{figure}
\setcounter{figure}{2}
\begin{figure}[t!]
\setlength{\unitlength}{1cm}
\begin{picture}(15,12)
\put(-1,0){\epsfig{file=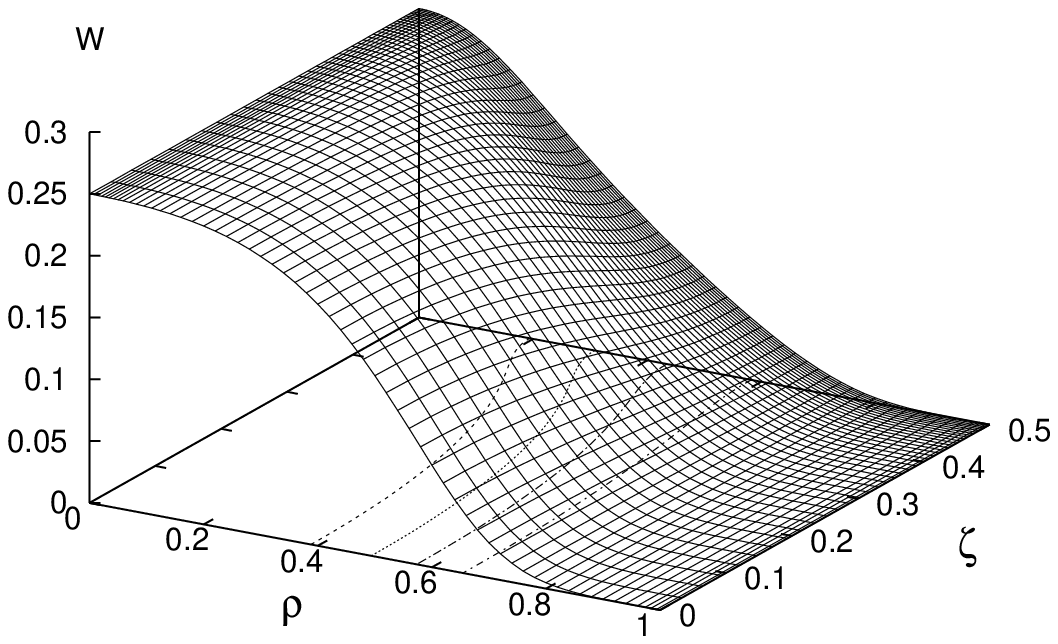,width=8cm}}
\put(7,0){\epsfig{file=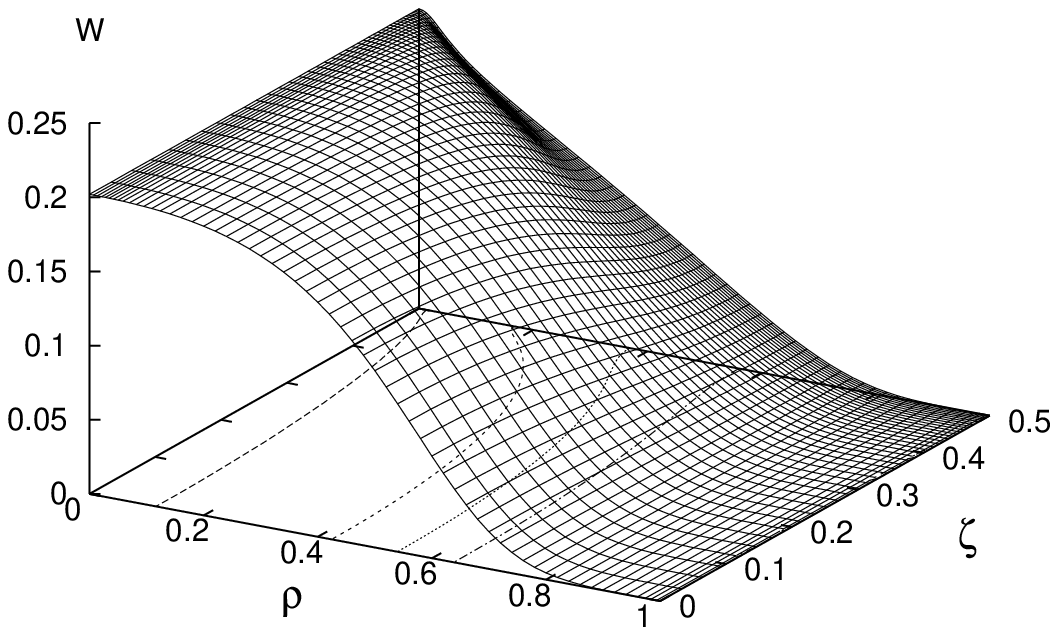,width=8cm}}
\put(-1,6){\epsfig{file=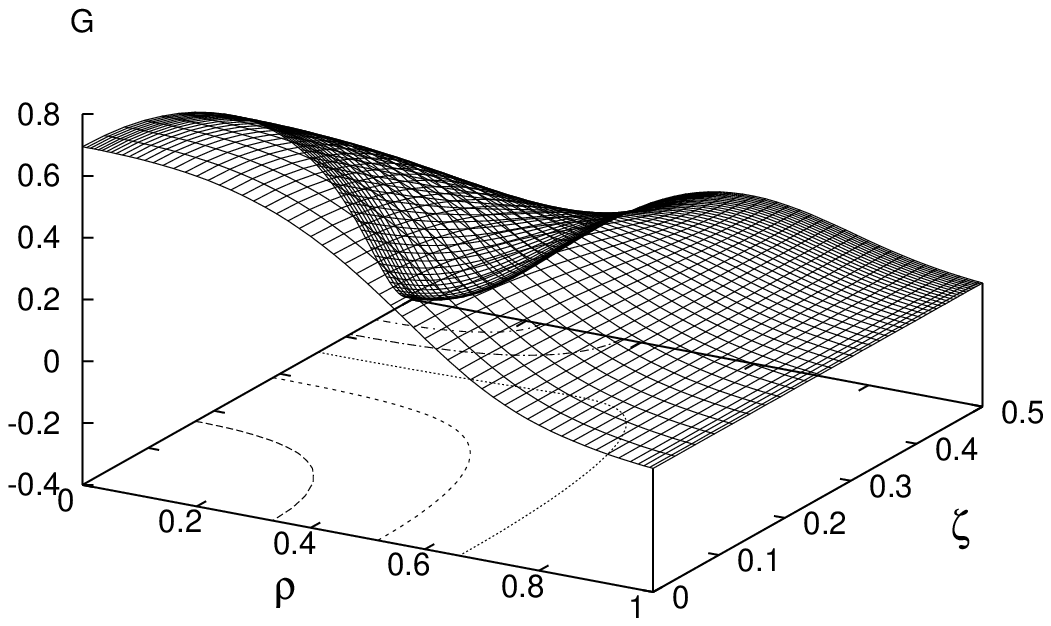,width=8cm}}
\put(7,6){\epsfig{file=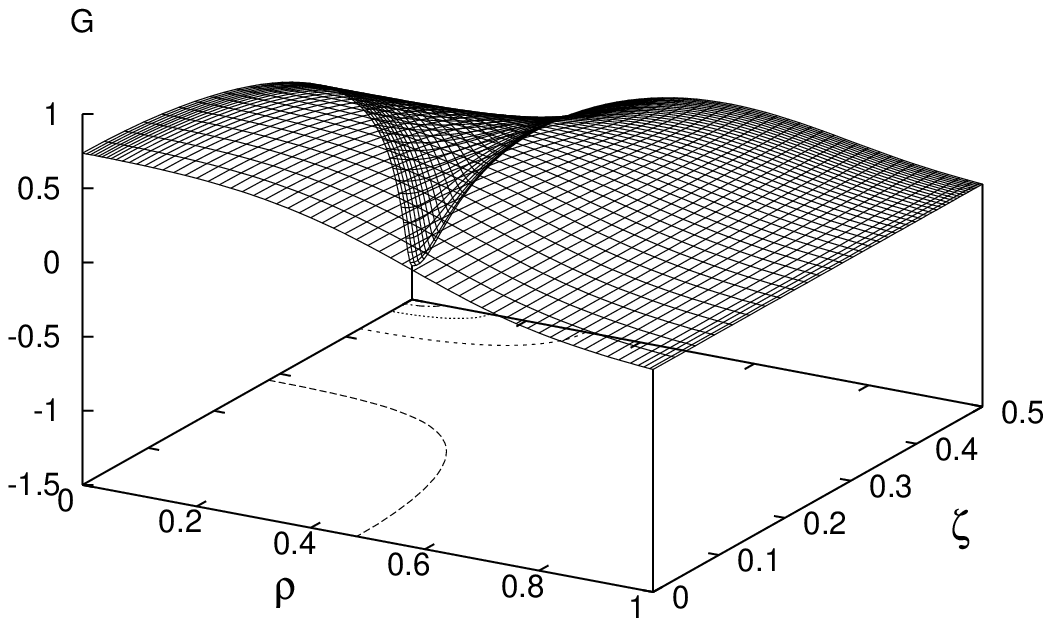,width=8cm}}
\end{picture}
\caption{
continued.
}
\end{figure}

\subsection{Properties of rotating black strings}
\subsubsection{The horizon}

For the static NUBS a measure of their deformation is given by the
nonuniformity parameter $\lambda$ \cite{Gubser:2001ac}
\begin{eqnarray} 
\lambda = \frac{1}{2} \left( \frac{{\cal R}_{\rm max}}{{\cal R}_{\rm min}}
 -1 \right)
, \label{lambda} 
\end{eqnarray}
where ${\cal R}_{\rm max}$ and ${\cal R}_{\rm min}$
represent the maximum radius of a $(D-3)$-sphere on the horizon and
the minimum radius, respectively,
the minimum radius being the radius of the `waist'
of the black string.
Thus for uniform black strings $\lambda=0$,
while the horizon topology
changing transition should be approached
for $\lambda \rightarrow \infty$ 
\cite{Kol:2003ja,Wiseman:2002ti}.
 
For the rotating NUBS one has to take into account, that
the rotation leads to a deformation of the 3-sphere of the horizon, making
it oblate w.r.t.~the planes of rotation.
Therefore, various possibilities arise to define 
the nonuniformity parameter $\lambda$.
In the following we employ the above definition of $\lambda$,
where ${\cal R}_{\rm max}$ and ${\cal R}_{\rm min}$
are obtained from the area $A_{H}$ of the respective
deformed 3-sphere via $A_{H}=2 \pi^2 {\cal R}^3$.

In Figure 4 we exhibit
the spatial embedding of the horizon into 3-dimensional space
for a sequence of $D=6$ rotating NUBS.
In these embeddings the symmetry directions ($\varphi_1$, $\varphi_2$)
are suppressed, and the proper circumference of the horizon is plotted
against the proper length along the compact direction,
yielding a geometrical view of both the deformation of 
the horizon due to rotation and the nonuniformity of the horizon
with respect to the compact coordinate.

For the solutions of the sequence shown in Figure 4
the temperature is kept fixed with temperature parameter $d_0=0.6$.
The first solution of the sequence 
corresponds to the marginally stable rotating uniform black string,
which has $\lambda=0$ and horizon angular velocity $\Omega_{H}=0.34908$.
When the horizon angular velocity is lowered,
rotating black strings with increasing nonuniformity are obtained.
Shown are solutions with nonuniformity parameter
$\lambda=0.83$, $1.7$ and $2.9$.
The latter is already close to the expected
topology changing transition to rotating caged black holes.
Interestingly, close to the maximal radius ${\cal R}_{\rm max}$
the deformation of the horizon due to rotation is significant,
whereas close to ${\cal R}_{\rm min}$ the 3-horizon appears
to be almost spherical.

\begin{figure}[t!]
\setlength{\unitlength}{1cm}
\begin{picture}(15,18)
\put(-4, 9){\epsfig{file=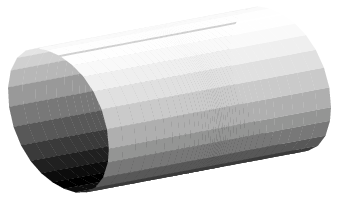,width=15cm}}
\put( 6.8,15.3){$\Omega_{H}=0.34908$}
\put( 6.7,14.7){$\lambda=0$}
\put( 4, 9){\epsfig{file=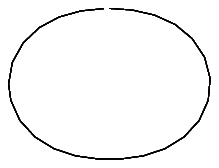,width=15cm}}
\put(-4, 5){\epsfig{file=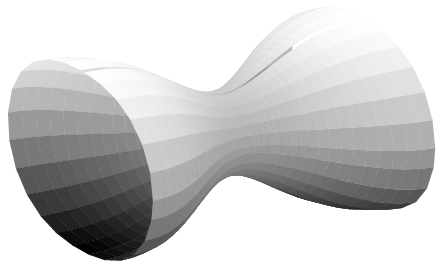,width=15cm}}
\put( 7.0,11.3){$\Omega_{H}=0.25$}
\put( 6.95,10.7){$\lambda=0.83$}
\put( 4, 5){\epsfig{file=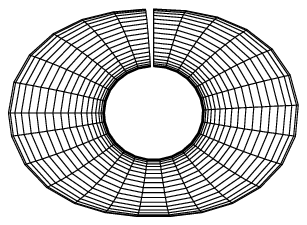,width=15cm}}
\put(-4, 1){\epsfig{file=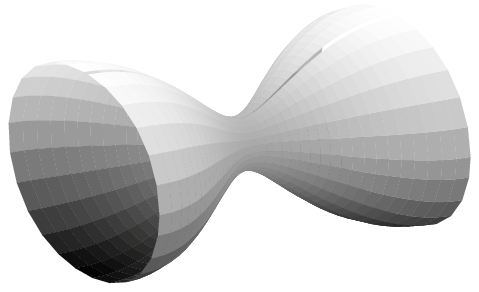,width=15cm}}
\put( 7.0, 7.3){$\Omega_{H}=0.212$}
\put( 6.95, 6.7){$\lambda=1.7$}
\put( 4, 1){\epsfig{file=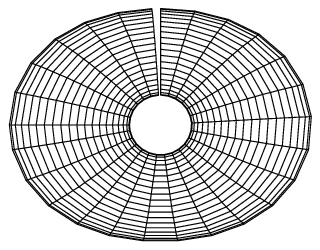,width=15cm}}
\put(-4,-3){\epsfig{file=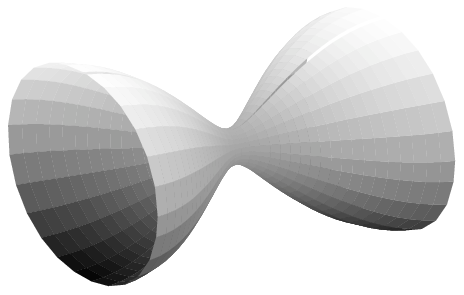,width=15cm}}
\put( 7.0, 3.3){$\Omega_{H}=0.202$}
\put( 6.95, 2.7){$\lambda=2.9$}
\put( 4,-3){\epsfig{file=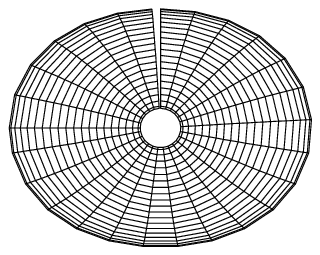,width=15cm}}

\end{picture}
\caption{
The spatial embedding of the horizon 
of $D=6$ rotating black string solutions
is shown for a sequence of solutions with fixed
temperature parameter $d_0=0.6$ and varying
horizon angular velocity $\Omega_{H}$:
$\Omega_{H}=0.34908$ (upper row), $\Omega_{H}=0.25$ (second row),
$\Omega_{H}=0.212$ (third row) and $\Omega_{H}=0.202$ (lower row),
$\lambda$ specifies the increasing nonuniformity of the solutions.
Left column: side view, right column: view in $z$ direction.
($r_0 =1$, $L=L^{\rm crit}=4.9516$.)
}
\end{figure}

The deformation of the horizon due to rotation 
is demonstrated in more detail in Figure 5,
where circumferences of the deformed 3-sphere
of the horizon are exhibited.
Here ${l}_{\rm e,max}$ denotes the equatorial maximum circumference,
and ${l}_{\rm e,min}$ the equatorial minimum circumference,
both referring to the circumferences in the two
equivalent planes of rotation, while
${l}_{\rm p,max}$ denotes the polar maximum circumference,
and ${l}_{\rm p,min}$ the polar minimum circumference,
representing circumferences for fixed azimuthal angles.
In the static case, the respective equatorial and polar circumferences
agree, and the minimum circumference represents the
circumference of the waist of the NUBS.

Using the scaled energy-mass $M_s$ and the scaled
angular momentum $J_s$, Eq.~(\ref{gen-defsc}),
we exhibit in Figure 5
these polar and equatorial circumferences 
versus the scaled angular momentum ratio $J_s/M_s^{3/2}$.
We note, that for rotating uniform black strings, this ratio is bounded,
$J_s/M_s^{3/2} \le 1$, with extremal rotating uniform solutions
saturating the bound.

For reference, the figure exhibits the polar and equatorial
circumferences of the branch of marginally stable MP 
uniform black strings.
This uniform branch ranges from the static marginally stable black string
to the extremal rotating marginally stable black string.
The static marginally stable string, with all circumferences equal,
has temperature parameter $d_0=0$.
Along this rotating UBS branch,
with equal maximal and minimal circumferences,
the temperature parameter $d_0$
and the deformation of the horizon 3-sphere due to rotation
both increase monotonically,
while the temperature itself decreases monotonically,
reaching zero in the extremal case.

\subsubsection{A critical temperature}
The rotating nonuniform black strings branch off from 
the marginally stable uniform black strings.
These branches are obtained,
by fixing a value of the temperature parameter $d_0$
and thus fixing the temperature,
and then decreasing the horizon angular velocity $\Omega_H$
from the respective rotating UBS value.
Depending on the value of the fixed chosen temperature parameter $d_0$,
the corresponding rotating NUBS branches exhibit distinct features.
When $d_0 < d_{0}^*$,
the rotating NUBS branch extends back to a static NUBS solution
with a finite waist, and thus finite minimal circumferences ${l}_{\rm p,min}$
and ${l}_{\rm e,min}$.
The size of the waist of the static NUBS solution
decreases with increasing $d_0$.
At the critical value $d_{0}^*$,
the respective branch of rotating NUBS is expected to extend precisely
back to a static solution with zero size waist,
i.e., to the solution at the topology changing transition,
where the branch of static NUBS merges with the branch
of static caged black holes.
This critical value of the temperature parameter $d_{0}^*$ 
is in the interval $0.30 < d_0 < 0.33$,
and corresponds to a critical value of the temperature $T_*$
where $0.72 < T_H/T_0 < 0.74$ (with $T_0$ the temperature
of the UBS).
This may be compared with our previous results
for static black strings \cite{Kleihaus:2006ee}.
Extraction of the critical temperature $T_*$ from
those static black string calculations,
suggests the bounds $0.72 < T_H/T_0 < 0.76$
for the critical temperature $T_*$ (when trying to account for
numerical inaccuracy in the critical region).\footnote{
Such an identification of $T_*$ assumes a monotonic
dependence of the temperature of the static uniform strings
on the nonuniformity parameter $\lambda$,
since otherwise bifurcations might be present
and complicate this scenario.}

Beyond the critical value $d_{0}^*$, the branches of rotating NUBS
no longer reach static NUBS.
Instead they are expected to extend to a
corresponding rotating solution with zero size waist,
and thus to lead towards a topology changing transition,
associated with the merging of a branch of rotating NUBS
and a branch of rotating caged black holes.
Indeed, when $d_0 > d_{0}^*$, 
the waist of the NUBS solutions monotonically decreases in size,
the minimal circumferences approaching zero.
Thus we see here first evidence, that 
a topology changing transition arises also for rotating 
branches of solutions.
We note, that the deformation of the horizon 3-sphere due to rotation
is considerable at maximum size, 
while the waist of the rotating NUBS becomes increasingly spherical.

\begin{figure}[t!]
\setlength{\unitlength}{1cm}
\begin{picture}(8,6)
\put(0,0.0){\epsfig{file=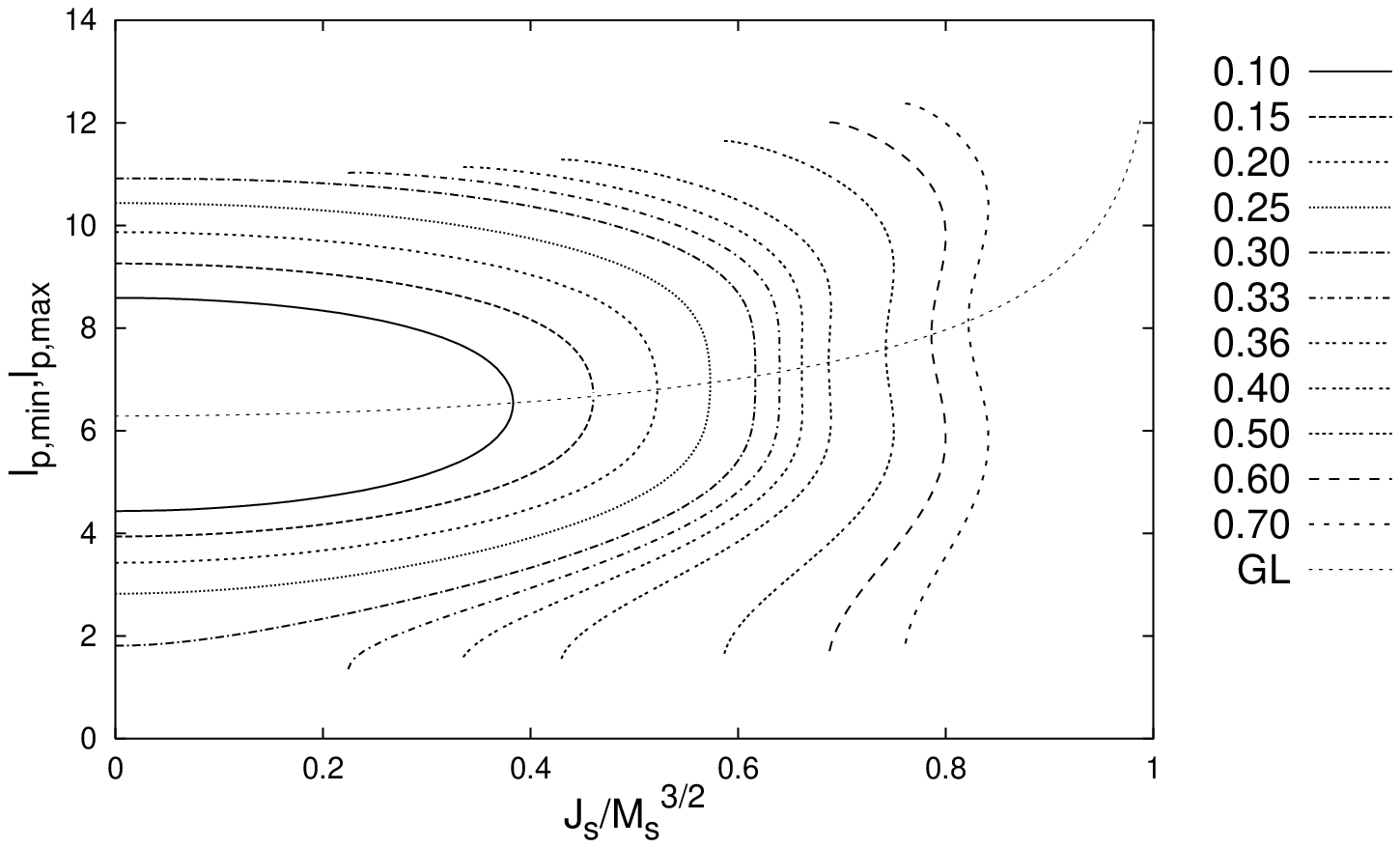,width=8.4cm}}
\put(8.5,0.0){\epsfig{file=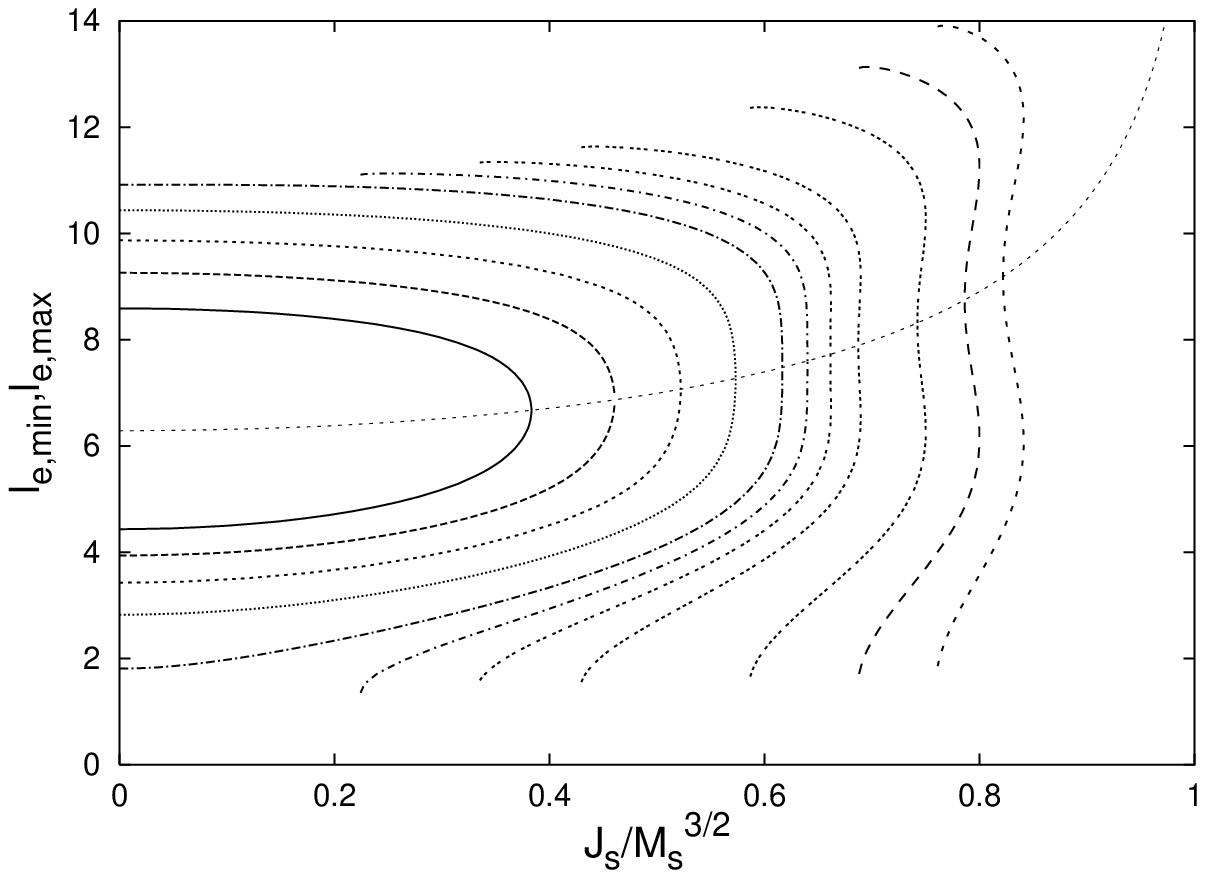,width=7.2cm}}
\end{picture}
\caption{
The maximum and minimum polar circumferences ${l}_{\rm p,max}$
and ${l}_{\rm p,min}$ of the deformed horizon $3$-sphere,
and the respective equatorial circumferences ${l}_{\rm e,max}$
and ${l}_{\rm e,min}$ are shown 
versus the scaled angular momentum ratio $J_s/M_s^{3/2}$
for branches of rotating NUBS with fixed values of the
temperature parameter $d_0$ ranging from 0.1 to 0.7.
The circumferences of the deformed horizon $3$-sphere
are also shown for marginally stable MP UBS (denoted GL).
}
\end{figure}

In Figure 6 we exhibit the nonuniformity parameter $\lambda$
versus the scaled angular momentum ratio $J_s/M_s^{3/2}$
for the same set of rotating NUBS. 
The branches begin at the rotating marginally stable UBS
with $\lambda=0$.
When $d_0 < d_{0}^*$,
the rotating NUBS branches extend back to static NUBS solutions
with finite nonuniformity and thus finite waist.
When $d_0 > d_{0}^*$, on the other hand,
the nonuniformity parameter $\lambda$ increases
apparently without bound,
approaching the topology changing transition for 
$\lambda \rightarrow \infty$.

The branch of rotating marginally stable UBS is bounded
by the static and by the extremal rotating solution.
It would be interesting to obtain the 
corresponding domain of existence of rotating NUBS.
The construction of extremal rotating NUBSs (if they exist), however,
currently represents an unsolved numerical challenge.

Figure 6 also exhibits the scaled horizon angular velocity
$\Omega_H/\Omega_{H,\rm GL}$ for this set of rotating NUBS
versus the scaled angular momentum ratio $J_s/M_s^{3/2}$.
Here $\Omega_{H,\rm GL}$ denotes the horizon angular velocity
of the marginally stable rotating UBS 
(with the same temperature parameter $d_0$).
Starting from rotating marginally stable UBS
with $\Omega_H/\Omega_{H,\rm GL}=1$, the branches
end at static NUBS with $\Omega_H=J=0$, when $d_0 < d_{0}^*$. 
When $d_0 > d_{0}^*$, in contrast,
the branches of rotating NUBS appear to approach limiting solutions 
with finite horizon angular velocity $\Omega_H$ and finite
angular momentum $J$, associated with 
a topology changing transition.

\begin{figure}[h!]
\setlength{\unitlength}{1cm}
\begin{picture}(8,6)
\put(0,0.0){\epsfig{file=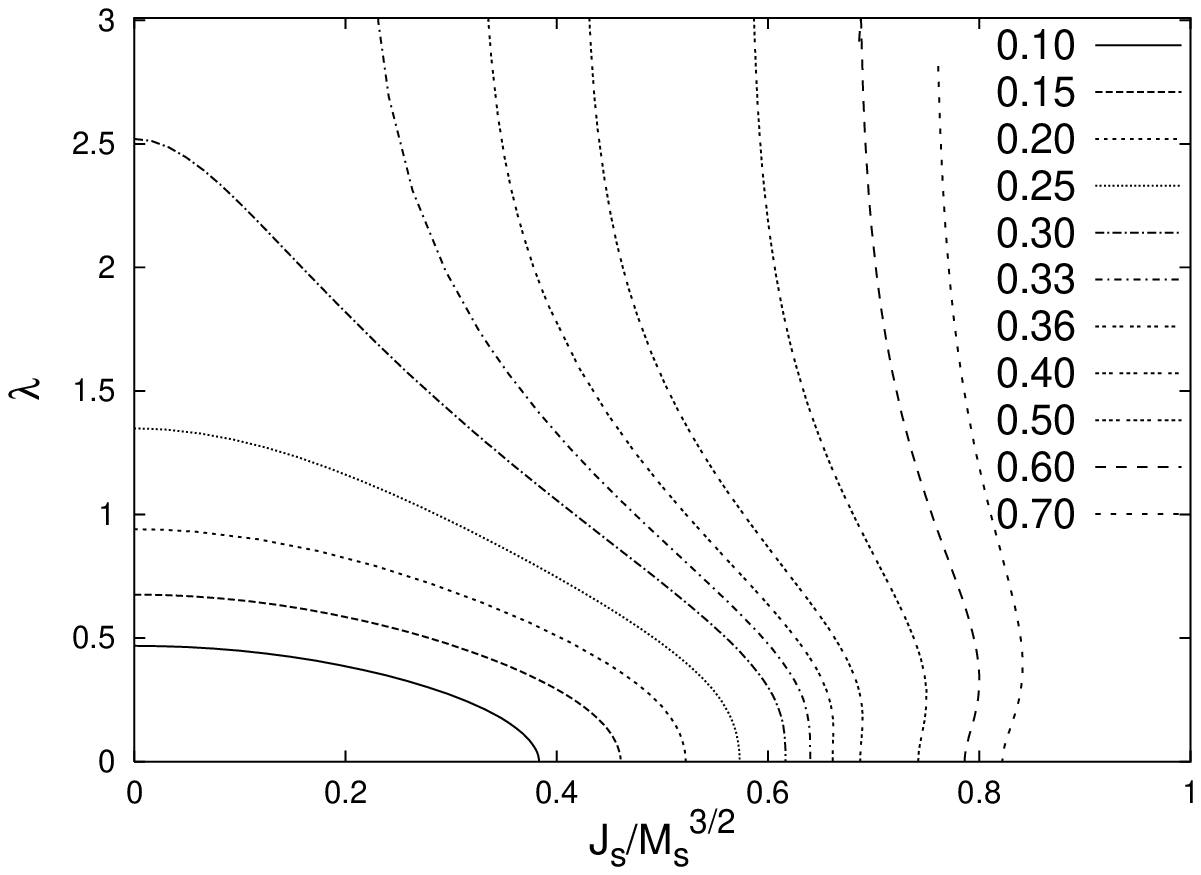,width=7.7cm}}
\put(8.0,0.0){\epsfig{file=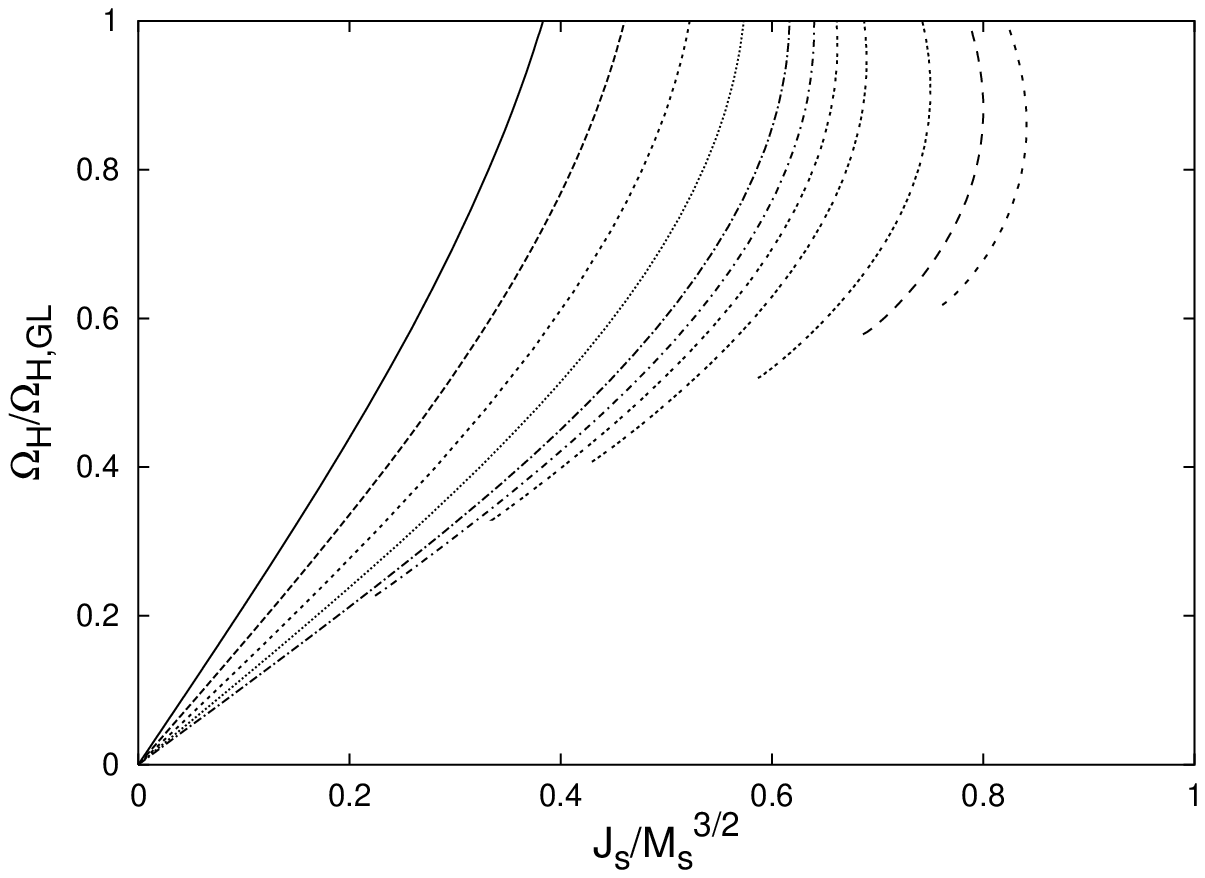,width=7.7cm}}
\end{picture}
\caption{
The nonuniformity parameter $\lambda$ 
and the scaled horizon angular velocity $\Omega_H/\Omega_{H,\rm GL}$ 
are shown
versus the scaled angular momentum ratio $J_s/M_s^{3/2}$
for branches of rotating NUBS with fixed values of the
temperature parameter $d_0$ ranging from 0.1 to 0.7.
($\Omega_{H,\rm GL}$ denotes the angular velocity
of the marginally stable MP UBS.)
}
\end{figure}
\subsubsection{Global charges}
The scaled mass $M_s/M_{s,\rm GL}$
and the scaled entropy $S_s/S_{s,\rm GL}$
(where $S= S_s {A_{D-3}L}/{4}$)
are exhibited in Figure 7
versus the scaled angular momentum ratio $J_s/M_s^{3/2}$
for the same set of solutions.
Both $M_s/M_{s,\rm GL}$ and $S_s/S_{s,\rm GL}$
increase monotonically along the branches of solutions
with fixed temperature.
As noted above, when $T_H>T_*$, 
the branches of rotating NUBS end at static NUBS,
whereas, when $T_H<T_*$, 
they appear to approach rotating limiting solutions
with finite values of the angular momentum $J$, associated with
a topology changing transition between branches of rotating solutions.
We conclude from the figure, that
the scaled mass $M_s/M_{s,\rm GL}$
of the limiting solutions increases with increasing $J_s/M_s^{3/2}$,
while their scaled entropy $S_s/S_{s,\rm GL}$
appears to be almost constant.

\begin{figure}[h!]
\setlength{\unitlength}{1cm}
\begin{picture}(8,6)
\put(0,0.0){\epsfig{file=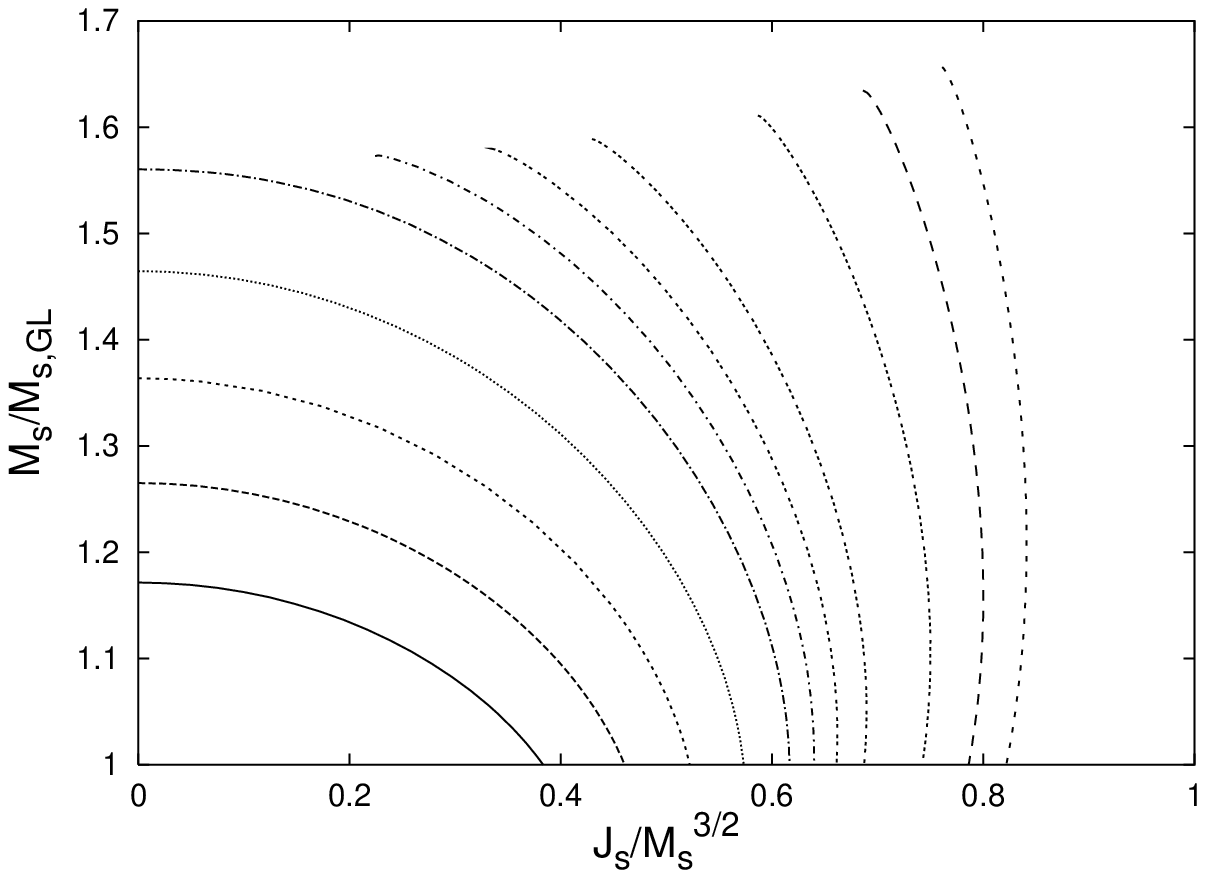,width=7.7cm}}
\put(8,0.0){\epsfig{file=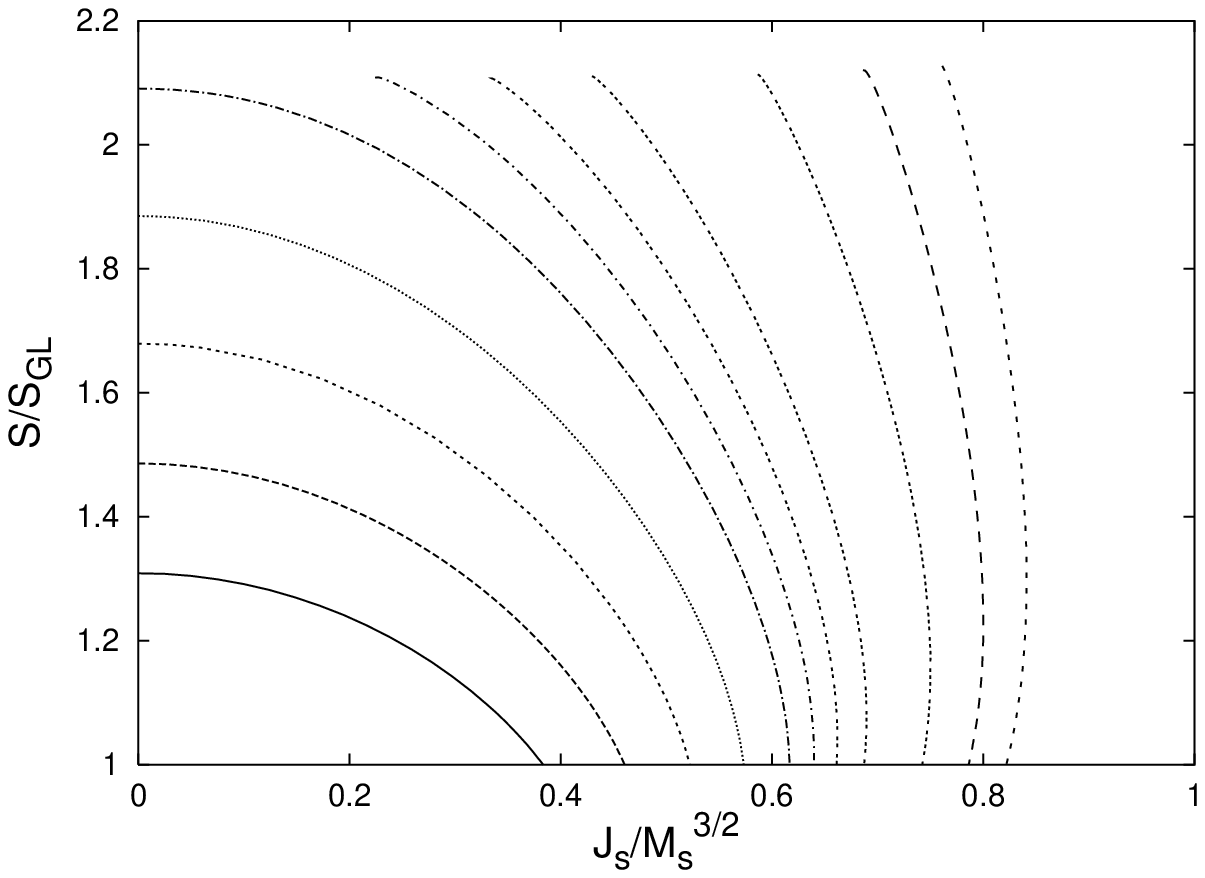,width=7.7cm}}
\end{picture}
\caption{
Same as Figure 6 for the scaled mass
$M_s/M_{s,\rm GL}$ and the scaled entropy $S_s/S_{s,\rm GL}$.
($M_{s,\rm GL}$ and $S_{s,\rm GL}$ denote the respective
quantities of the marginally stable MP UBS.)
}
\end{figure}

We exhibit the relative tension $n$ of this set of rotating NUBS
in Figure 8, together with the relative
tension of the uniform black strings, $n_{\rm GL}=1/3$.
Starting from rotating marginally stable UBS,
the relative tension $n$ decreases monotonically
for branches of rotating NUBS with large values of the temperature.
As the critical temperature $T_*$ is approached,
and beyond the critical temperature,
the tension $n$ no longer decreases monotonically,
but instead reaches a minimum and then increases again.
Thus we observe the backbending phenomenon for the relative tension $n$,
encountered previously for static NUBS \cite{Kleihaus:2006ee},
also for rotating nonuniform black strings.
For the static NUBS we obtained for the relative tension $n$
the critical value $n_* \approx 0.2$.
Consistency requires, that this value
agrees within error bounds with the critical value obtained here
for the branch of rotating NUBS at the critical temperature $T_*$.
The figure indicates, that this requirement may hold.

\begin{figure}[h!]
\setlength{\unitlength}{1cm}
\begin{picture}(8,6)
\put(3.7,0.0){\epsfig{file=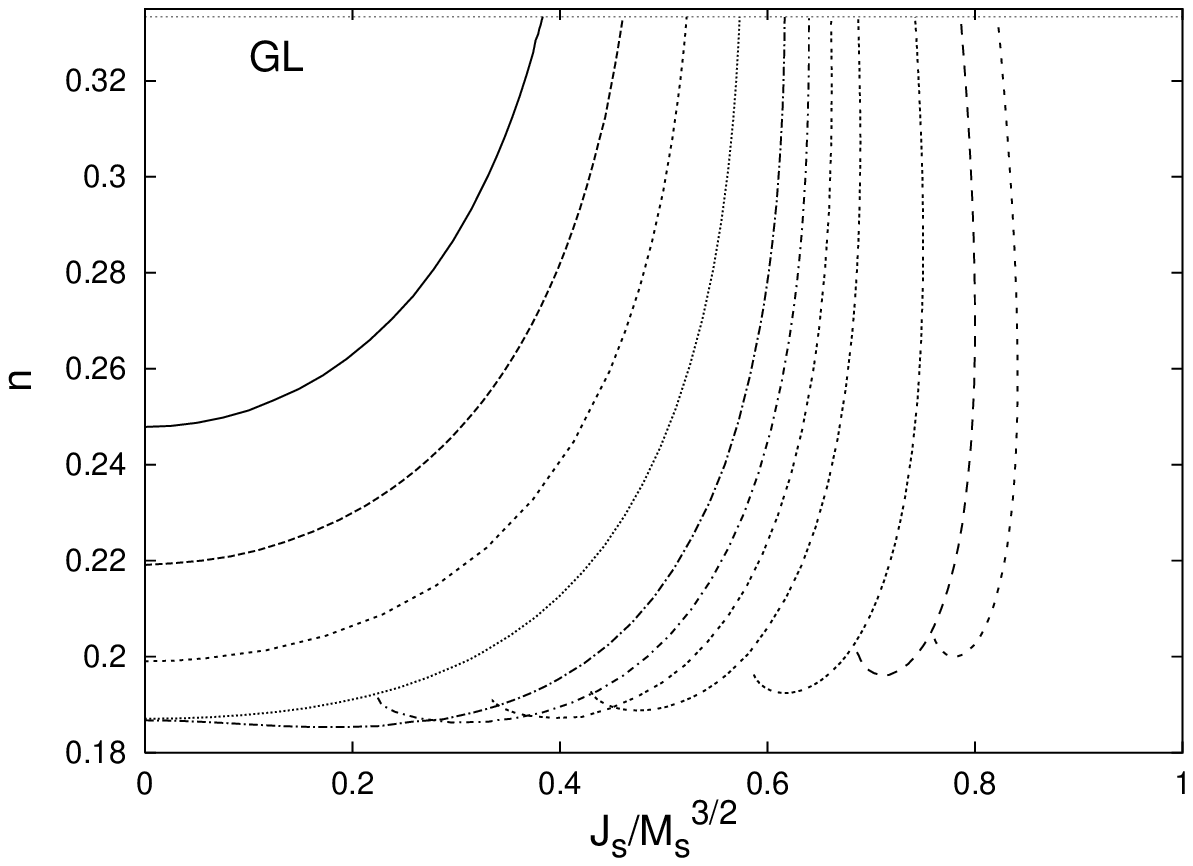,width=7.7cm}}
\end{picture}
\caption{
Same as Figure 6 for the relative tension $n$.
}
\end{figure}
Restricting to a canonical ensemble, the numerical analysis
indicates that the qualitative thermodynamical features of the
uniform MP branch are also shared by rotating NUBS solutions.
For small values of $J$, the entropy is a decreasing function of $T$,
i.e. $C_{J,L}<0$. However, the configurations near extremality are
thermally stable in a canonical ensemble.
Also, although further work is necessary in this
case, we expect all nonuniform solutions to be thermodynamically
unstable in a
grand canonical ensemble.
\subsubsection{The ergoregion}
Like rotating black holes,
rotating black strings possess an ergoregion.
While the ergosurface of rotating uniform black strings
is uniform like the horizon, the ergosurface of
nonuniform black strings reflects the nonuniformity
of the horizon.
In Figure 9 we exhibit
the spatial embedding of the ergosurface into 3-dimensional space
for those rotating NUBS, whose
spatial embedding of the horizon was shown in Figure 4.

\begin{figure}[h!]
\setlength{\unitlength}{1cm}
\begin{picture}(15,18)
\put(-4, 9){\epsfig{file=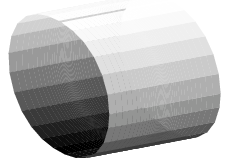,width=15cm}}
\put( 6.8,15.3){$\Omega_{H}=0.34908$}
\put( 6.7,14.7){$\lambda=0$}
\put( 4, 9){\epsfig{file=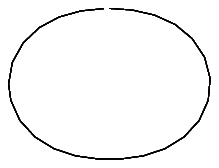,width=15cm}}
\put(-4, 5){\epsfig{file=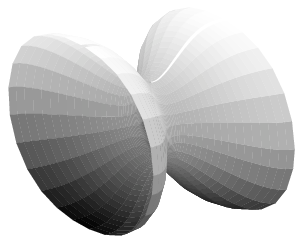,width=15cm}}
\put( 7.0,11.3){$\Omega_{H}=0.25$}
\put( 6.95,10.7){$\lambda=0.83$}
\put( 4, 5){\epsfig{file=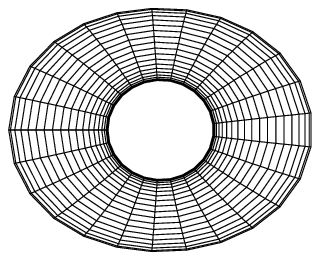,width=15cm}}
\put(-4, 1){\epsfig{file=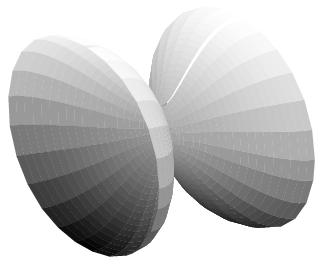,width=15cm}}
\put( 7.0, 7.3){$\Omega_{H}=0.212$}
\put( 6.95, 6.7){$\lambda=1.7$}
\put( 4, 1){\epsfig{file=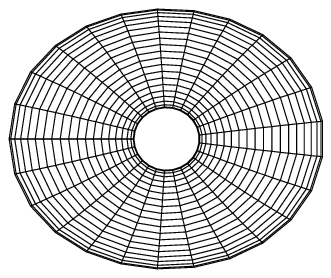,width=15cm}}
\put(-4,-3){\epsfig{file=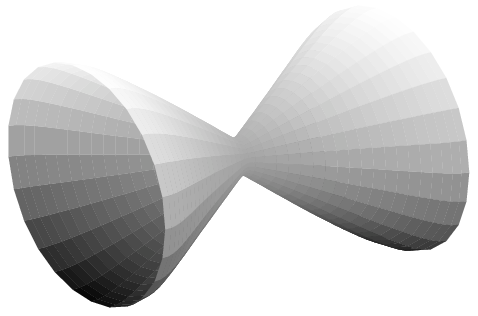,width=15cm}}
\put( 7.0, 3.3){$\Omega_{H}=0.202$}
\put( 6.95, 2.7){$\lambda=2.9$}
\put( 4,-3){\epsfig{file=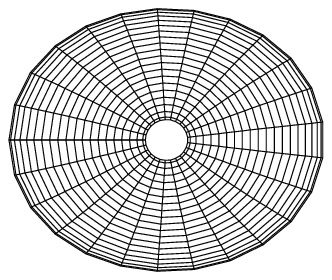,width=15cm}}
\end{picture}
\caption{
The spatial embedding of the ergosurface.
of $D=6$ rotating black string solutions
is shown for a sequence of solutions with fixed
temperature parameter $d_0=0.6$ and varying
horizon angular velocity $\Omega_{H}$:
$\Omega_{H}=0.34908$ (upper row), $\Omega_{H}=0.25$ (second row),
$\Omega_{H}=0.212$ (third row) and $\Omega_{H}=0.202$ (lower row),
$\lambda$ specifies the increasing nonuniformity of the solutions.
Left column: side view, right column: view in $z$ direction.
($r_0 =1$, $L=L^{\rm crit}=4.9516$.)
}
\end{figure}

As in Figure 4, the symmetry directions are suppressed here.
The proper circumference of the ergosurface is plotted
against the proper arclength along the compact direction.
Denoting $\tilde{r}_e(z)$, $z$ the coordinates of the ergosurface at
fixed $\varphi_1$,  $\varphi_2$, $\theta$, $t$,
we define the arclength as
$$\sigma =
\int_0^z e^{B(\tilde{r}_e(z),z)}
\sqrt{1+\left(\frac{d\tilde{r}_e}{dz}\right)^2}dz.
$$

The solutions shown have fixed temperature
with temperature parameter $d_0=0.6$,
decreasing horizon angular velocity $\Omega_{H}$,
and increasing nonuniformity of the horizon.
The first solution
corresponds to the marginally stable rotating uniform black string,
with an ergosurface uniform w.r.t.~the compact coordinate,
but rotational deformation.
When the horizon angular velocity is lowered,
the nonuniformity of the ergosurface increases, along with the
increase of the nonuniformity of the horizon.

In Figure 10 we exhibit the ergosurface in terms of the
coordinates $r_{\rm ergo} = \tilde r$ and $\zeta$,
in which the horizon is located at $\tilde r=0$.
We show rotating NUBS solutions on three branches with fixed temperature,
corresponding to the values of the temperature parameter 
$d_0=0.3$, $0.33$ and $0.6$.
Thus the first two sets represent solutions just above
and just below the critical temperature $T_*$.
All sets start with the corresponding rotating UBS 
and thus a uniform ergosurface.
As the horizon angular velocity $\Omega_H$ decreases from $\Omega_{H,\rm GL}$
nonuniformity of the ergosurface develops.

\begin{figure}[h!]
\setlength{\unitlength}{1cm}
\begin{picture}(8,12)
\put(0,6.0){\epsfig{file=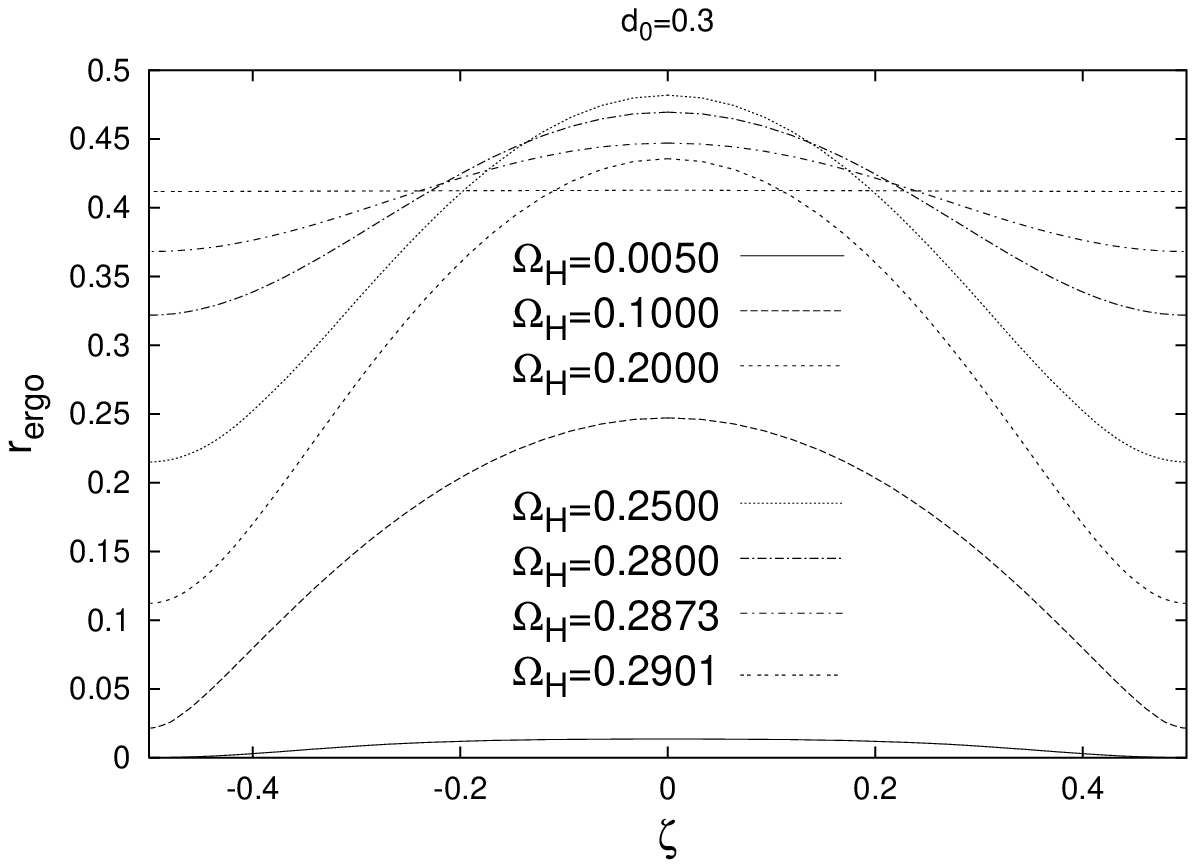,width=7.7cm}}
\put(8,6.0){\epsfig{file=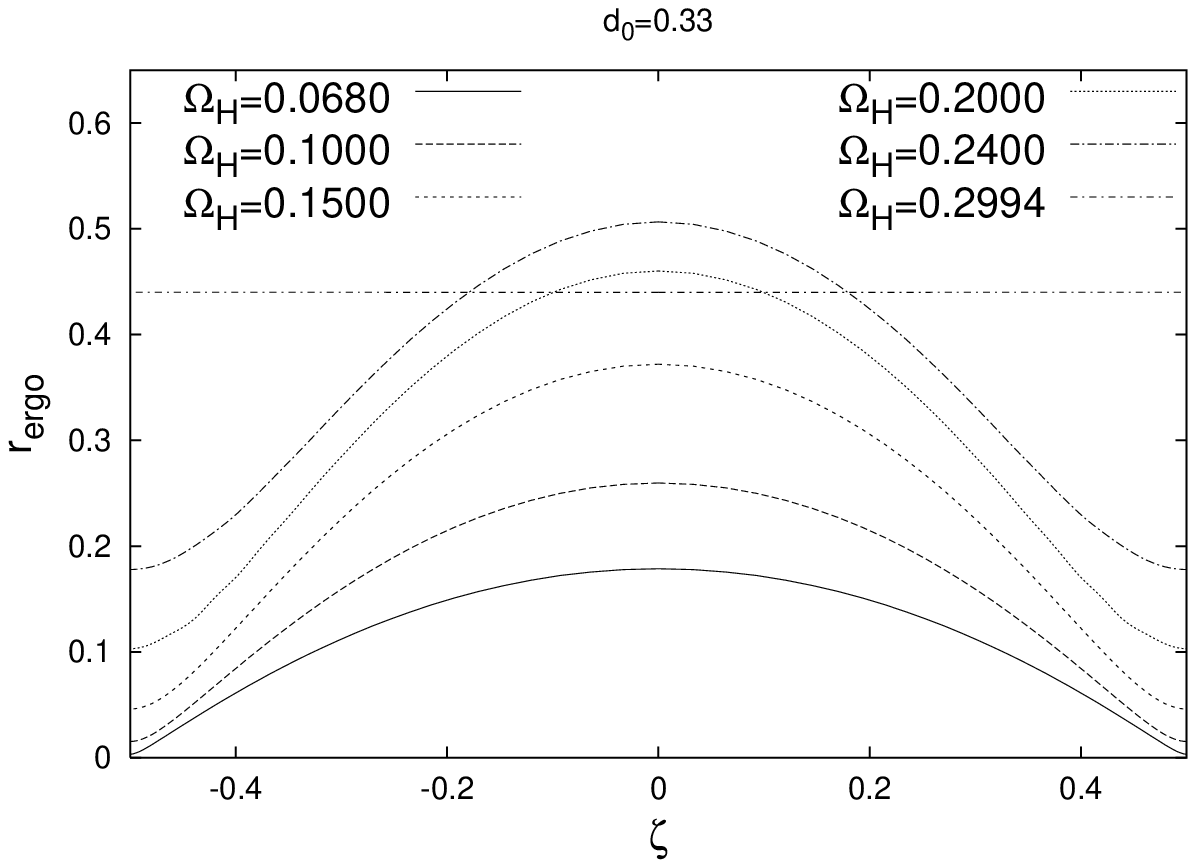,width=7.7cm}}
\put(3.7, 0.0){\epsfig{file=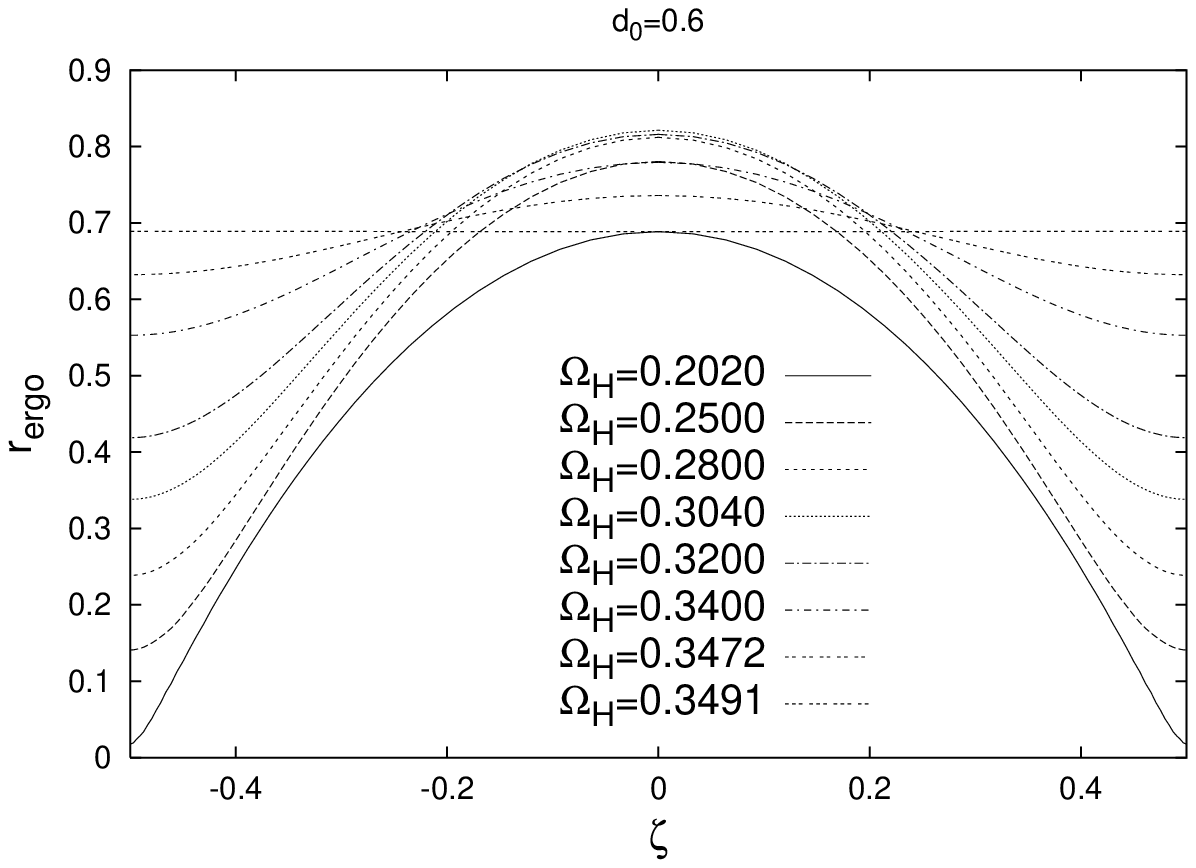,width=7.7cm}}
\end{picture}
\caption{
The ergosurface of rotating NUBS is shown
for three sets of solutions with fixed temperature,
corresponding to the values of the temperature parameter,
$d_0=0.3$, $0.33$ and $0.6$,
and decreasing horizon angular velocity $\Omega_H$, starting from
the value $\Omega_{H,\rm GL}$ 
of the respective marginally stable MP UBS.
}
\end{figure}

For $d_0=0.3$, where $T_H > T_*$,
the rotating NUBS branch extends back to a static NUBS solution,
where the ergoregion disappears. The strong shrinkage of the ergoregion
close to this point is clearly seen in the figure for the NUBS solution 
with horizon angular velocity $\Omega_H=0.005$.
For $d_0=0.33$, the temperature is just below the critical value,
$T_H < T_*$, and the rotating NUBS branch is expected to extend to a
corresponding rotating solution with zero size waist,
and thus signify a topology changing transition
between rotating solutions.
This is reflected in the figure by the presence of a finite ergoregion
of the solution with $\Omega_H=0.068$, close to the transition point.
Also for $d_0=0.6$, the ergoregion remains finite, as the 
topology changing transition is approached.
We note, that the size of the ergoregion of the limiting
solution appears to increase with decreasing temperature.

\section{Rotating NUBS in heterotic string theory }

In order to obtain  rotating electrically charged black strings,
we employ a solution generating technique, by performing symmetry transformations on
the neutral solution. Within toroidally compactified heterotic string, an approach
to obtain the charged solutions from the neutral one was presented in Ref.
 \cite{Sen:1994eb}.
This method was used to obtain, e.g., general rotating electrically charged solutions in four
dimensions
 \cite{Sen:1994eb}, 
higher-dimensional general electrically charged static solutions \cite{Peet:1995pe},
and rotating black hole
solutions with one rotational parameter in $D$ dimensions \cite{Horowitz:1995tm}.
General $D-$dimensional charged rotating black holes
 with $[(D-1)/2]$ distinct angular momenta
were constructed in \cite{Cvetic:1996dt}.



The massless fields in heterotic string theory compactified on a 
$(10-D)$-dimensional torus consist of the string
metric $G_{\mu\nu}$, the anti-symmetric
tensor field $B_{\mu\nu}$, $(36-2D)$ U(1) gauge fields $A_\mu^\ab$
($1\le j\le 36-2D$), the scalar dilaton field $\Phi$, and a
$(36-2D)\times (36-2D)$
matrix valued scalar field $M$ satisfying,
\begin{eqnarray} 
 \label{e5}
M L M^T = L, \quad \quad M^T=M.
\end{eqnarray}
Here $L$ is a $(36-2D)\times(36-2D)$
symmetric matrix with $(26-D)$ eigenvalues $-1$ and
$(10-D)$ eigenvalues $+1$. Following \cite{Sen:1994eb}, \cite{Peet:1995pe} 
we shall take $L$ to be
\begin{eqnarray}  \label{e4}
L=\pmatrix{-I_{26-D} & \cr & I_{10-D}\cr},
\end{eqnarray}
where $I_n$ denotes an $n\times n$ identity matrix. The action describing
the effective field theory of these massless
bosonic fields is given by \cite{Maharana:1992my}
\begin{eqnarray}  
\label{gen-act-string}
S &=&  \int d^D x \sqrt{-\det G} \, e^{-\Phi} \, \Big[ R_G + G^{\mu\nu}
\p_\mu \Phi \p_\nu\Phi +{1\over 8} G^{\mu\nu} Tr(\p_\mu M L\p_\nu ML)
\nonumber \\
&& -{1\over 12} G^{\mu\mu'} G^{\nu\nu'} G^{\rho\rho'} H_{\mu\nu\rho}
H_{\mu'\nu'\rho'} - G^{\mu\mu'} G^{\nu\nu'} F^\ab_{\mu\nu} \, (LML)_{jk}
\, F^\bb_{\mu'\nu'} \Big] \, ,
\end{eqnarray}
where
\begin{eqnarray}  
\label{enq2}
F^\ab_{\mu\nu} = \p_\mu A^\ab_\nu - \p_\nu A^\ab_\mu \, ,
\end{eqnarray}
\begin{eqnarray} 
\label{enq3}
H_{\mu\nu\rho} = \p_\mu B_{\nu\rho} + 2 A_\mu^\ab L_{jk} F^\bb_{\nu\rho}
+\hbox{cyclic permutations of $\mu$, $\nu$, $\rho$}\, ,
\end{eqnarray}
and $R_G$ denotes the scalar curvature associated with the metric
$G_{\mu\nu}$. The canonical Einstein metric $\bar{g}_{\mu \nu}$ is
\begin{equation}
\label{stringeinstein}
\bar g_{\mu\nu} = e^{-\frac{2}{D-2}\Phi}G_{\mu\nu} .
\end{equation}

One can add a general charge to any stationary vacuum solution  by
applying the solution generating transformations $(O(26-D,1)/O(26-D))
\times (O(10-D,1)/O(10-D))$. 
 This generates a nontrivial $\Phi, B_{\mu\nu}$ and
$M$, as well as $A_\mu^\ab$. 
 
Here we are mainly interested in the case $D=6$,
i.e., heterotic string theory compactified on a four-dimensional
torus.
 Since the generation procedure is
identical to the one given in Ref.\cite{Sen:1994eb} we shall not give the
details, but only present the final results. 

Starting with an arbitrary time independent
solution $g_{\mu \nu}$
of the vacuum field equation in $D=6$ (with $ds^2=g_{\mu \nu}dx^\mu dx^\nu$),
the expression for the metric of a
charged configuration expressed in the canonical Einstein frame is
\begin{eqnarray} 
\label{new1}
 d\bar s^2=\Delta^{1/4} \bigg(ds^2+\frac{1}{g_{tt}}
(\frac{(\cosh \alpha+\cosh \beta)^2}{4 \Delta}-1)
(g_{\varphi_1t}d\varphi_1+g_{\varphi_2 t}d\varphi_2)^2
\\
\nonumber
+(\frac{ \cosh \alpha+\cosh \beta }{\Delta}-2)
(g_{\varphi_1t}d\varphi_1dt+g_{\varphi_2 t}d\varphi_2dt)
+(\frac{1-\Delta}{\Delta})g_{tt}dt^2\bigg)~,
\end{eqnarray}
where $\Delta$ is related to the dilaton,
\begin{eqnarray} 
\label{new2}
e^{-2\Phi}= \Delta= \frac{1}{4}\bigg(g_{tt}^2(\cosh \alpha-\cosh \beta)^2
+2g_{tt}(\sinh^2 \alpha+\sinh^2 \beta)
+(\cosh\alpha+\cosh\beta)^2
\bigg).
\end{eqnarray}
The time components of the $U(1)$ gauge fields are
\begin{eqnarray}
\nonumber 
 A_t^{(i)} &=&
-\frac{n^{(i)}}{4\sqrt{2}\Delta}(1+g_{tt})\sinh \alpha 
\bigg(\cosh \alpha+\cosh \beta
+g_{tt}(\cosh \alpha-\cosh \beta)\bigg),
~~1\leq i \leq 20,
\\
\nonumber
&=&-\frac{p^{(i-20)}}{4\sqrt{2}\Delta}(1+
g_{tt})\sinh \beta
\bigg(\cosh \alpha+\cosh \beta
-g_{tt}(\cosh \alpha-\cosh \beta)\bigg),
~~21\leq i \leq  24,
\end{eqnarray}
whereas the spatial components of the gauge fields are given by
\begin{eqnarray}
\nonumber
 A_{\varphi_k}^{(i)} &=& 
\frac{n^{(i)}g_{t\varphi_k  }}{2\sqrt{2}g_{tt}}
\sinh\alpha \bigg(
1-\frac{(1+g_{tt})}{4\Delta} (\cosh \alpha+\cosh \beta)
(\cosh \alpha+\cosh \beta
+g_{tt}(\cosh \alpha-\cosh \beta))\bigg),
\\
\nonumber
&&~~~~~~~~~~~~~~~~~~~~~~~~~~~~~~~~~~~~~~~ 
~~~~~~~~~~~~~~~~~~~~~~~~~~~~~~~~~~~~~~~~~~~~~~~~~
~~1\leq i \leq 20,
\\
\nonumber
&=&
\frac{p^{(i-20)}g_{t\varphi_k  }}{2\sqrt{2}g_{tt}}
\sinh\beta \bigg(
1-\frac{(1+g_{tt})}{4\Delta} (\cosh \alpha+\cosh \beta)
(\cosh \alpha+\cosh \beta-g_{tt}(\cosh \alpha-\cosh \beta))\bigg),
\\
\nonumber
&&~~~~~~~~~~~~~~~~~~~~~~~~~~~~~~~~~~~~~~~ 
~~~~~~~~~~~~~~~~~~~~~~~~~~~~~~~~~~~~~~~~~~~~~~~~~
~~21\leq i \leq 24,
\end{eqnarray}
with $k=1,2$.
Here $\alpha$ and $\beta$ are two boost angles,
$\vec n$ is a 20-dimensional unit vector, and $\vec p$ is a 4-dimensional unit
vector.
The nonvanishing components of the two-form field $B_{\mu \nu}$ are
\begin{eqnarray} 
\label{enq5}
\hspace{-0.5cm}
B_{t \varphi_k} =
\frac{g_{t \varphi_k}}{2g_{tt}}
(\cosh \alpha-\cosh \beta)\bigg(1-\frac{(1+g_{tt})}{4\Delta} 
 ((\sinh^2 \alpha+\sinh^2 \beta)g_{tt}
+(\cosh \alpha+\cosh\beta)^2)\bigg).
\end{eqnarray}
The result for the matrix-valued scalar $M$ is
\begin{eqnarray} 
\label{M-matrix}
M = I_{24} +\pmatrix{ P_1~nn^T & P_2~n p^T \cr P_2~p n^T & P_3~pp^T\cr}\, ,
\end{eqnarray}
where
\begin{eqnarray} 
\nonumber
P_1&=&\frac{\sinh^2 \alpha~(1+g_{tt})^2}{2g_{tt}}
(1-\frac{1}{4\Delta}\left(\cosh \alpha+\cosh \beta +
g_{tt}(\cosh \alpha-\cosh \beta))^2\right) ,
\\
P_2&=&-\frac{2\sinh \alpha \sinh \beta~(1+g_{tt})}{4\Delta}
\left(1+\cosh \alpha \cosh \beta +g_{tt}
(\cosh \alpha\cosh \beta-1) \right),
\\
\nonumber
P_3&=&\frac{\sinh^2 \beta~(1+g_{tt})^2}{2g_{tt}}
(1-\frac{1}{4\Delta}\left(\cosh \alpha+\cosh \beta -
g_{tt}(\cosh \alpha-\cosh \beta))^2\right). 
\end{eqnarray}
Both 
 black hole and black string  charged rotating
solutions can
be generated in this way.
The results in \cite{Sen:1994eb}
are recovered
for the case of a $D=6$ Myers-Perry black hole
solution
with only one nonzero angular momenta.

Charged rotating NUBS strings with two equal angular momenta
 are found by replacing in 
(\ref{new1})-(\ref{M-matrix}) the seed metric $ds^2$
as given by
$(\ref{metric})$
with
$g_{tt}=e^{2G}r^2W^2-e^{2A}(1-r_0^2/r^2)$,~~
$g_{\varphi_1 t}=-e^{2G}r^2 W \sin^2 \theta$,  
$g_{\varphi_2 t}=-e^{2G}r^2 W \cos^2 \theta$.
Here we present only the line element of the
 charged rotating UBS solution
\begin{eqnarray} 
d\bar s^2&&=\Delta(r)^{\frac{1}{4}}
\left(\frac{dr^2}{1-\frac{2M}{r^2}+\frac{2a^2M}{r^4}}+dz^2
+r^2(d \theta^2+\sin^2\theta d \varphi_1^2+\cos^2\theta d \varphi_2^2)\right)
\\
\nonumber
&&+\Delta(r)^{-\frac{3}{4}}\bigg(
 \frac{2aM}{r^2(1-\frac{2M}{r^2})}
(\Delta(r)-\frac{M}{2r^2}(\cosh \alpha+\cosh \beta))
(\sin^2\theta d \varphi_1 +\cos^2\theta d \varphi_2)^2
\\
\nonumber
&&-(\cosh \alpha+\cosh \beta)\frac{2aM}{ r^2}
(\sin^2\theta d \varphi_1dt +\cos^2\theta d \varphi_2dt)
-(1-\frac{2M}{r^2})dt^2 \bigg),
\end{eqnarray}
where
\begin{eqnarray} 
\Delta(r)=1+\frac{2M}{r^2}(\cosh \alpha \cosh \beta-1)
+\frac{M^2}{r^4}(\cosh \alpha-\cosh \beta)^2.
\end{eqnarray}
The extremal limit is found by taking the boost parameters
to infinity  together with
a rescaling of $M$.

The asymptotic structure of a general charged solution is
similar to that of the vacuum seed configuration.
The spacetime  still approaches the
${\cal M}^{5}\times S^1$ background as $r \to \infty $.
Also, one can see that the event horizon location of the charged solutions is unchanged.
The position of the ergosurface remains the same,
all fields being well defined on that hypersurface.
One can also verify that no closed timelike curves are introduced
in the line element (\ref{new1}) by the generation procedure. 

The relevant properties of a charged solution  can be derived from the
corresponding  vacuum seed configuration.
The mass-energy $\bar{E}$ and the string tension $\bar{\mathcal T}$
of the charged rotating solutions are
\begin{eqnarray} 
\label{new-charges}
\bar{E}= \frac{1}{4} (1+3\cosh \alpha \cosh \beta) E
+\frac{1}{4} (1-\cosh \alpha \cosh \beta){\mathcal T}_0L, 
~~~
\bar{\mathcal T}={\mathcal T}.
\end{eqnarray}
The charged solution possesses also two equal angular momenta
$\bar J_1=\bar J_2=\bar{J}$ with
\begin{eqnarray} 
 \bar{J}=\frac{1}{2}(\cosh \alpha+\cosh \beta)J.
\end{eqnarray}
and has event horizon velocities  $
\bar{\Omega}_1=\bar{\Omega}_2=
\bar{\Omega}_H$ with
\begin{eqnarray} 
\bar{\Omega}_H=\frac{2\Omega_H}{\cosh \alpha+\cosh \beta}.
\end{eqnarray}
The electric charges defined as
\begin{eqnarray} 
\label{Qe-def}
Q_e^{(k)}=\frac{1}{LA_3}\lim_{r \to \infty}\int dz \int dA_3 r^3F_{rt}^{(k)}
\end{eqnarray}
have the following expression
\begin{eqnarray} 
\label{Qe}
 Q_e^{(i)}&=&\frac{G_6}{\sqrt{2}\pi}\ 
(\frac{3E}{L}-{\mathcal T})n^{(i)}\sinh \alpha \cosh \beta ~,~~~~~~~1\leq i \leq 20,
\\
\nonumber
 &=&\frac{G_6}{\sqrt{2}\pi}\ 
(\frac{3E}{L}-{\mathcal T})p^{(i-20)}\sinh \beta  \cosh \alpha~,~~~21\leq i \leq 24.
\end{eqnarray}
The new solutions have a nonzero magnetic moment,
\begin{eqnarray} 
\label{mu}
 \mu^{(i)}&=&\frac{G_6\sqrt{2} }{ \pi L}J n^{(i)}\sinh \alpha ~,~~~~~~~1\leq i \leq 20,
\\
\nonumber
 &=&\frac{G_6\sqrt{2} }{ \pi L}J p^{(i-20)}\sinh \beta~,~~~21\leq i \leq 24.
\end{eqnarray}
Their dilaton charge  
is
\begin{eqnarray} 
\label{Qd-def}
Q_d =\frac{G_6}{2\pi}\ 
(\frac{3E}{L}-{\mathcal T}) (\cosh \alpha \cosh \beta-1).
\end{eqnarray}
The relations between the Hawking temperature $\bar{T}_H$
and the entropy $\bar{S}$ of the charged solutions
and the corresponding quantities $T_H$ and $S$ of the vacuum seed solution are
\begin{eqnarray} 
\bar{T}_H=\frac{2T_H}{\cosh \alpha+\cosh \beta},~~
\bar{S}= \frac{1}{2}S(\cosh \alpha+\cosh \beta).
\end{eqnarray}
One can see that the entropy increases with the increase of
angular momentum.
On the other hand, the temperature decreases with 
the increase of the angular momenta.
However, the products of temperature and entropy, 
$\bar{T}_H \bar{S}$,
and of horizon angular velocity and angular momentum, 
$\bar{\Omega}_H \bar J$,
are independent of the
parameters $\alpha$, $\beta$.

As they should, all these properties reduce to those
of the neutral solution upon sending the 
boost parameters $\alpha$, $ \beta$ to zero.

We conclude, that every vacuum solution is associated
with a family of charged solutions, which depends on 24 free parameters.
In particular, the branch of non-uniform solutions emerging from
the uniform black string at the threshold unstable mode thus must
persist for strings with non-zero electric charge.

\section{Further remarks. Conclusions. }

Considering rotating black strings in $D$ dimensions,
we have first addressed the GL instability of MP UBSs with equal
magnitude angular momenta in even spacetime dimensions,
taking advantage of the enhanced symmetry of these configurations.
Expanding around the UBS and solving the eigenvalue problem numerically,
our results indicate that the GL
instability persists for these solutions up to extremality
for all even dimensions between six and fourteen.
This agrees with GM correlated stability conjecture
\cite{Gubser:2000ec}, since these
black objects are also  thermodynamically unstable
  in a grand canonical ensemble.
It may be interesting to note that the  GM conjecture
was also confirmed in \cite{Miyamoto:2006nd} for the case of
static, magnetically charged black strings,
the Gregory-Laflamme mode vanishing at the
point where the UBS becomes thermodynamically stable 
(which in that case is away from extremality).

While for static vacuum black strings study of the perturbative equations
in second order revealed the appearance of a critical dimension,
above which the perturbative nonuniform black strings
are less massive than the marginally stable uniform black string
\cite{Sorkin:2004qq},
the analogous study in the presence of rotation has yet to be
achieved.

In $D=6$ we then constructed numerically 
rotating nonuniform black strings with equal angular momenta.
These emerge from the branch of 
marginally stable rotating MP UBS solutions,
which ranges from the static marginally stable black string
to the extremal rotating marginally stable black string.
Along this UBS branch, the Hawking temperature $T_H$ decreases monotonically,
reaching zero in the extremal limit.
Fixing the value of the temperature (or equivalently
the temperature parameter) and decreasing the
value of the horizon angular velocity from the GL value,
then yields a corresponding branch of rotating nonuniform black strings.

Previously, in $D=6$ dimensions, evidence was provided
that the branch of static nonuniform black strings
and the branch of static caged black holes merge at a
topology changing transition \cite{Kudoh:2004hs},
the transition occurring at critical values of
the temperature $T_*$, the string tension $n_*$, etc.
The results we have found for  rotating NUBS
indicate that at $T_*$ the branches of rotating NUBS, each with fixed temperature,
exhibit another critical phenomenon. 

In particular, the branches of rotating NUBS at fixed temperature $T_H>T_*$
end at static NUBS solutions
with finite nonuniformity and thus finite waist,
where the nonuniformity of these static solutions increases 
as $T_H$ is decreasing towards $T_*$.
In contrast, along the branches of rotating NUBS at fixed temperature $T_H<T_*$
the nonuniformity parameter $\lambda$ increases
apparently without bound,
while at the same time the horizon angular velocity
and the angular momentum appear to approach finite values.
Thus for $T_H<T_*$ we see first evidence
for a topology changing transition,
where - in analogy with the static case -
branches of rotating nonuniform black strings
and branches of rotating caged black holes are expected to merge.

We conjecture that there exists a whole branch of rotating
singular topology changing solutions, labelled by their
decreasing temperature,
beginning with the static solution at $T_*$,
and leading possibly up to an extremal rotating solution at $T_H=0$.
Obtaining the respective branches of rotating caged black holes
represents a major numerical challenge.

In principle, there is yet another class of black objects which may 
play a role in this picture.
Apart from configurations with an $S^3\times S^1$ (black strings) and $S^4$ (black holes)
topology of the event horizon, the $D=6$ KK theory
 possesses also vacuum uniform solutions 
with an event horizon of topology $S^2\times S^1\times S^1$,
corresponding to uplifted $D=5$ black rings \cite{Emparan:2001wn,Pomeransky:2006bd}.
Nonuniform solutions with an event horizon of topology $S^2\times S^1\times S^1$,
approaching at infinity the ${\cal M}^{5}\times S^1$ background are also 
likely to exist.
They may join the MP NUBS branch at a topology changing transition.
However, for the case discussed in this paper with two equal 
magnitude angular momenta, we could not find any indication
of this scenario.
This appears to be consistent with the recent results in \cite{Pomeransky:2006bd},
where the general $D=5$ black  ring solution with two angular momenta was presented.
An inspection of this solution indicates that black rings with 
equal angular momenta must exhibit some pathologies,
which may explain our result.

We remark that, as in the static case \cite{Kleihaus:2006ee},
we observe the backbending phenomenon for the relative tension $n$
also for branches of rotating nonuniform black strings, 
below some critical value of the temperature.

Our last concern was the construction
of charged rotating NUBS in heterotic string theory,
by adding charge to the vacuum solutions by
applying solution generating $(O(26-D,1)/O(26-D))
\times (O(10-D,1)/O(10-D))$ transformations
\cite{Sen:1994eb}.
The properties of these new configurations 
can be derived from the corresponding vacuum solutions.

We expect that, similar to the
static case \cite{Harmark:2007md},  the solutions discussed in this paper may be relevant for the 
thermal phase structure of non-gravitational theories,  via gauge/gravity duality.

The construction of the general rotating 
vacuum NUMBS with distinct angular
momenta seems to represent an exceedingly difficult task.
However, the case of only one nonvanishing angular momentum
appears to be treatable.
These solutions may be found by using similar techniques to those
employed in this work and are currently under study. 

Although the static higher-dimensional black holes are stable
\cite{Ishibashi:2003ap},
their rotating counterparts need not be, at least for large rotation.
Recently, the existence of an effective Kerr bound
for $d>4$ rapidly rotating black holes with one nonzero angular momentum
was conjectured by Emparan and Myers \cite{Emparan:2003sy}. They showed
that the geometry of the event horizon of such rapidly rotating black objects 
in six or higher dimensions behaves like a black membrane.
Therefore the black hole becomes unstable.
This instability should persist for the corresponding rotating UBS solutions,
and is not associated with the extra dimensions.

Rotating black objects extending in extra dimensions
may also exhibit other instabilities.
Cardoso and Lemos \cite{Cardoso:2004zz} 
uncovered a new universal instability
for rotating black branes and strings, which holds for any
massless field perturbation. The main point of their argument
is that transverse dimensions in a black string geometry
act as an effective mass for the fields, which simulates a
mirror enclosing a rotating black hole, thereby creating
a black hole bomb.
For further work on the  instabilities of rotating black objects, see for example 
\cite{Gibbons:2002pq}, \cite{Marolf:2004fy}, \cite{Cardoso:2006ks}.

\section*{Acknowledgements}
 
B.K.~gratefully acknowledges support by the DFG under contract
KU612/9-1.
The work of E.R. was carried out in the framework of Enterprise--Ireland
Basic Science Research Project SC/2003/390.
 

\end{document}